%% file: main.tex
\documentclass[sigconf]{acmart}
\AtBeginDocument{%
  \providecommand\BibTeX{{%
    \normalfont B\kern-0.5em{\scshape i\kern-0.25em b}\kern-0.8em\TeX}}}

\copyrightyear{2024}
\acmYear{2024}
\setcopyright{acmlicensed}\acmConference[CHI '24]{Proceedings of the CHI Conference on Human Factors in Computing Systems}{May 11--16, 2024}{Honolulu, HI, USA}
\acmBooktitle{Proceedings of the CHI Conference on Human Factors in Computing Systems (CHI '24), May 11--16, 2024, Honolulu, HI, USA}
\acmDOI{10.1145/3613904.3642589}
\acmISBN{979-8-4007-0330-0/24/05}

\usepackage{lipsum}
\usepackage{caption}
\usepackage{subcaption}
\usepackage{enumitem}

\newcommand{\userquote}[1]{\textit{``#1''}}
\newcommand{\revision}[1]{#1}
\newcommand{\camready}[1]{\textcolor[rgb]{0, 0, 0}{#1}}

\definecolor{strong guidance}{HTML}{7F2704}
\definecolor{weak guidance}{HTML}{00441B}
\definecolor{adaptive guidance}{HTML}{3F007D}
\definecolor{no guidance}{HTML}{08306B}

\begin{document}

\title[FocusFlow: 3D Gaze-Depth Interaction in Virtual Reality]{FocusFlow: 3D Gaze-Depth Interaction in Virtual Reality Leveraging Active Visual Depth Manipulation}

\author{Chenyang Zhang}
\authornote{Both authors contributed equally to this research.}
\email{zhang414@illinois.edu}
\orcid{0009-0003-1116-4895}
\affiliation{%
  \institution{University of Illinois at Urbana-Champaign}
  \streetaddress{201 North Goodwin Avenue MC 258}
  \city{Urbana}
  \state{Illinois}
  \country{USA}
}

\author{Tiansu Chen}
\authornotemark[1]
\email{tiansuc2@illinois.edu}
\orcid{0009-0007-7227-7948}
\affiliation{%
  \institution{University of Illinois at Urbana-Champaign}
  \streetaddress{201 North Goodwin Avenue MC 258}
  \city{Urbana}
  \state{Illinois}
  \country{USA}
}

\author{Eric Shaffer}
\email{shaffer1@illinois.edu}
\orcid{0009-0007-8656-7225}
\affiliation{%
  \institution{University of Illinois at Urbana-Champaign}
  \streetaddress{201 North Goodwin Avenue MC 258}
  \city{Urbana}
  \state{Illinois}
  \country{USA}
}

\author{Elahe Soltanaghei}
\email{elahe@illinois.edu}
\orcid{0009-0006-5040-5438}
\affiliation{%
  \institution{University of Illinois at Urbana-Champaign}
  \streetaddress{201 North Goodwin Avenue MC 258}
  \city{Urbana}
  \state{Illinois}
  \country{USA}
}

\renewcommand{\shortauthors}{Zhang and Chen, et al.}

\begin{abstract}
Gaze interaction presents a promising avenue in Virtual Reality (VR) due to its intuitive and efficient user experience. Yet, the depth control inherent in our visual system remains underutilized in current methods. In this study, we introduce FocusFlow, a hands-free interaction method that capitalizes on human visual depth perception within the 3D scenes of Virtual Reality. We first develop a binocular visual depth detection algorithm to understand eye input characteristics. We then propose a layer-based user interface and introduce the concept of ``Virtual Window'' that offers an intuitive and robust gaze-depth VR interaction, despite the constraints of visual depth accuracy and precision spatially at further distances. \revision{Finally, to help novice users actively manipulate their visual depth, we propose two learning strategies that use different visual cues to help users master visual depth control. Our user studies on 24 participants demonstrate the usability of our proposed virtual window concept as a gaze-depth interaction method. In addition, our findings reveal that the user experience can be enhanced through an effective learning process with adaptive visual cues, helping users to develop muscle memory for this brand-new input mechanism.} We conclude the paper by discussing potential future research topics of gaze-depth interaction.
\end{abstract}



\begin{teaserfigure}
 \centering
 \begin{subfigure}[b]{0.32\textwidth}
     \centering
     \includegraphics[width=\textwidth]{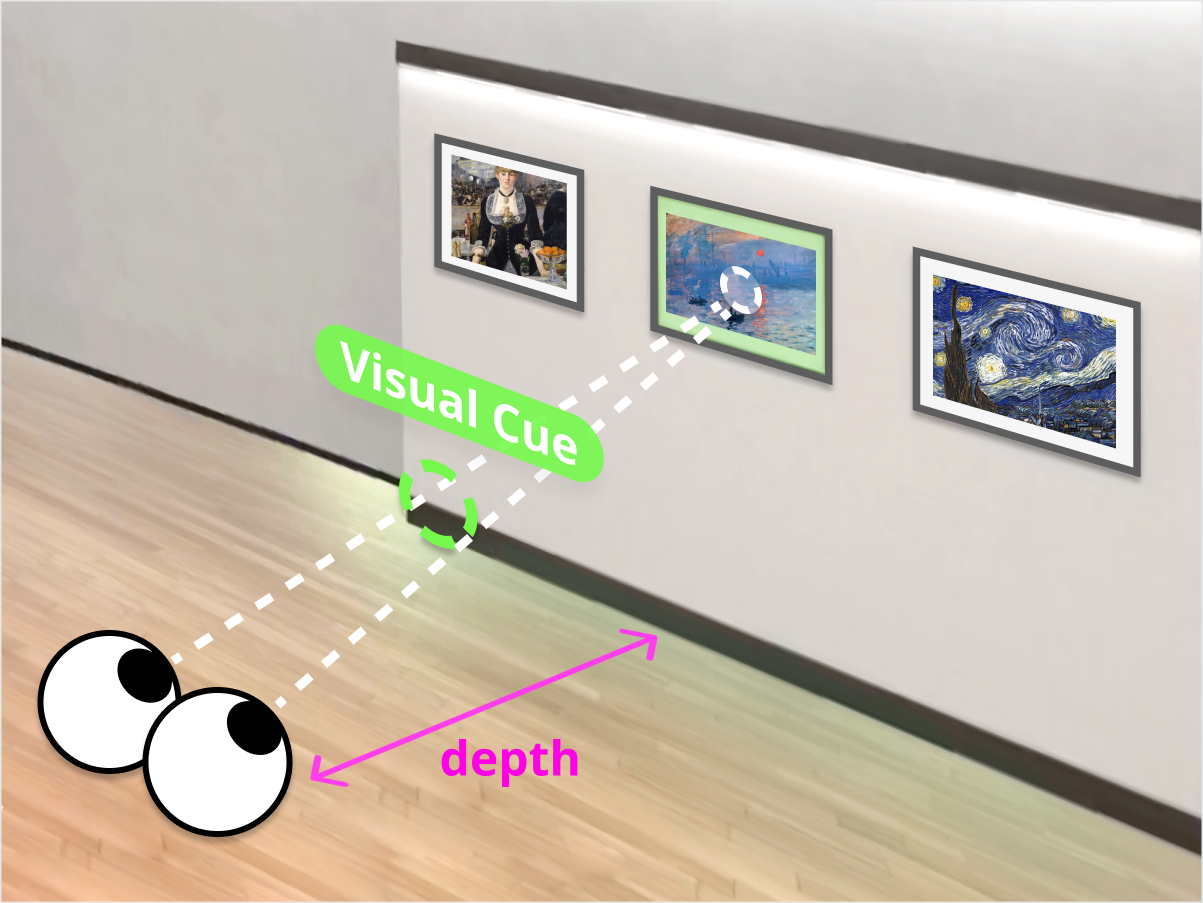}
     \caption{Hovering Over Target Object}
     \label{fig:teaser-a}
 \end{subfigure}
 \hfill
 \begin{subfigure}[b]{0.32\textwidth}
     \centering
     \includegraphics[width=\textwidth]{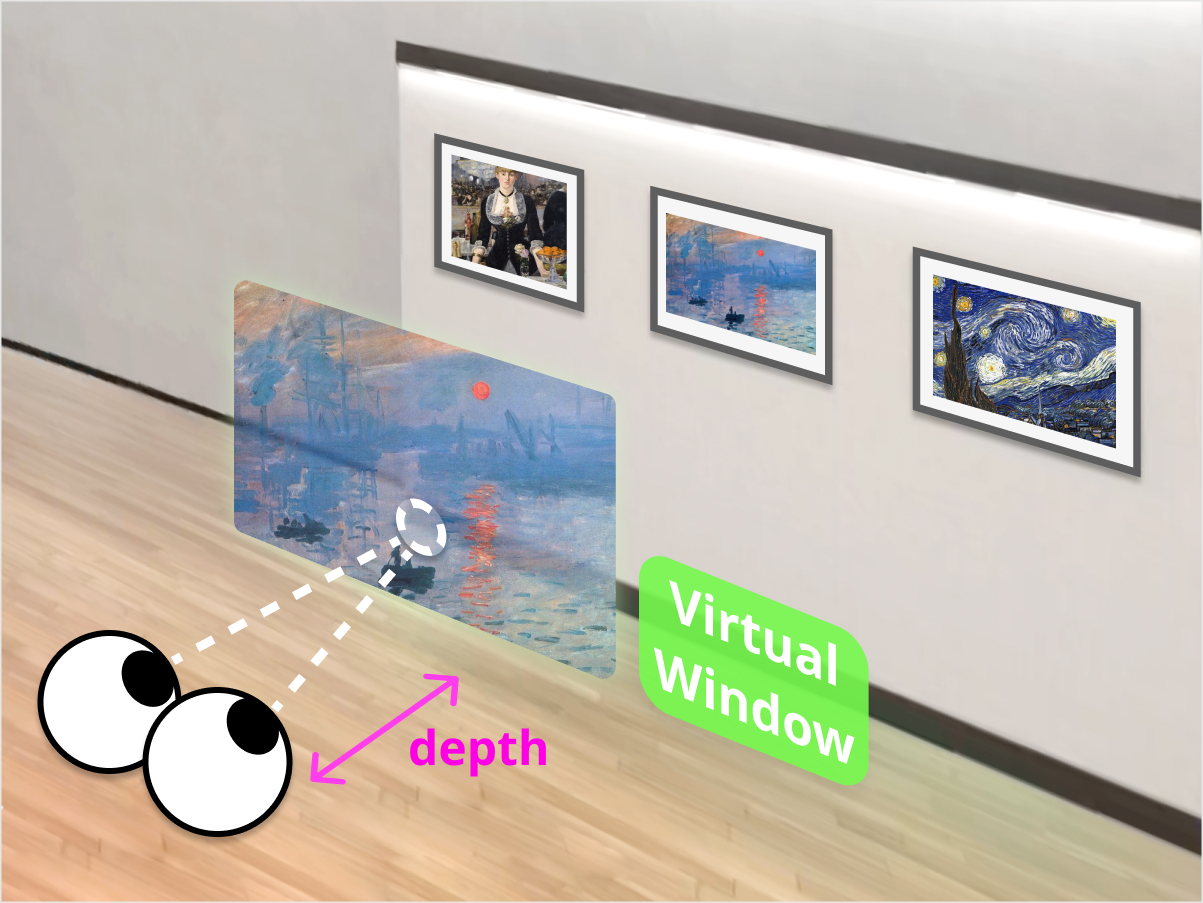}
     \caption{Activation Through Pulling Closer}
     \label{fig:teaser-b}
 \end{subfigure}
 \hfill
 \begin{subfigure}[b]{0.32\textwidth}
     \centering
     \includegraphics[width=\textwidth]{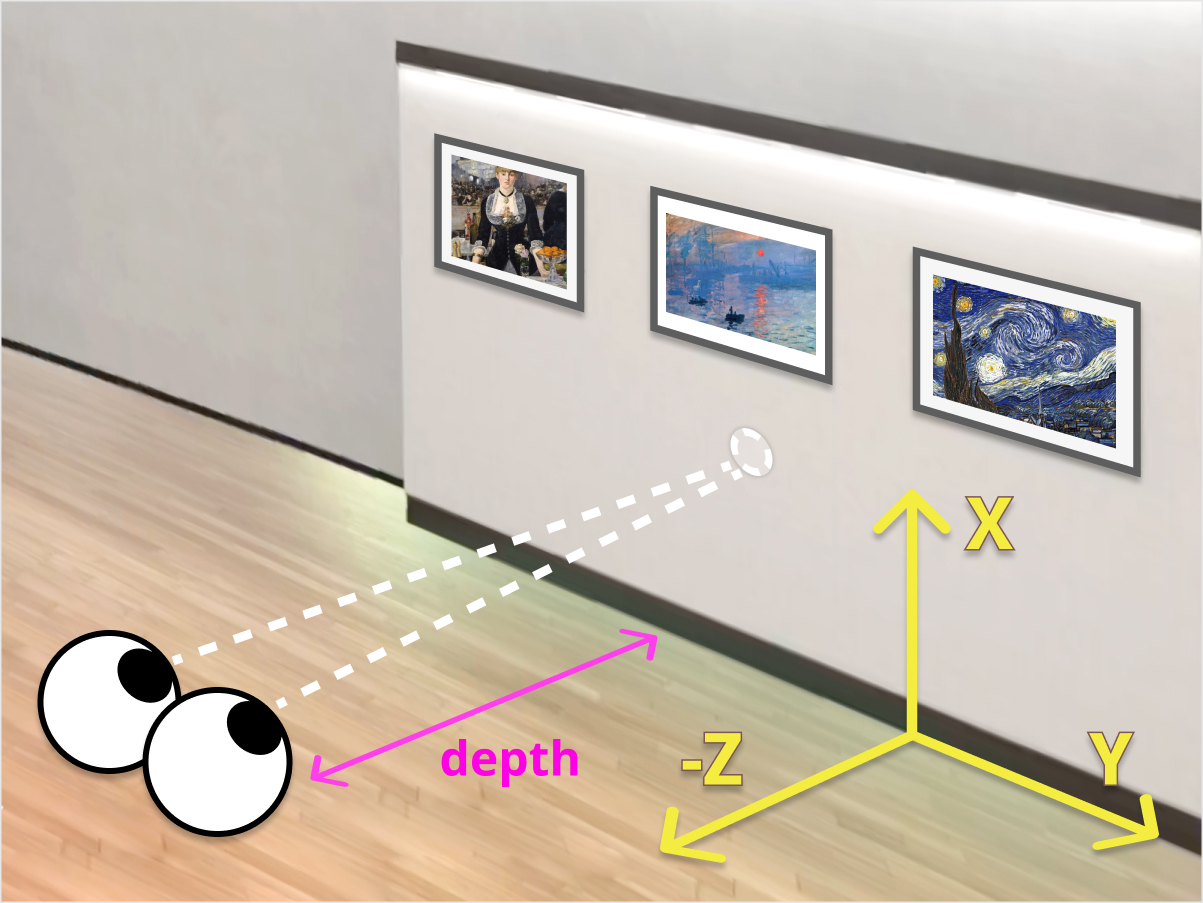}
     \caption{Exit Through Pushing Further}
     \label{fig:teaser-c}
 \end{subfigure}
\caption{Concept of FocusFlow. (a) When users hover over an object, a visual cue will appear to guide the users to shift their visual depth between layers at different distances. (b) Users can perform hands-free selection to activate a \emph{Virtual Window} in front, by pulling their visual depth closer along the z-dimension. (c) Users can exit the \emph{Virtual Window} by naturally pushing the visual depth further back to the wall.}
\label{fig:teaser}
\Description{FocusFlow is a gaze interaction method that enhances traditional methods by incorporating the parallax of the user's eyes to calculate visual depth. By seamlessly integrating z-dimension information into the interaction logic, users can intuitively shift their gaze point between different layers, enabling hands-free selection.}
\end{teaserfigure}


\maketitle

\section{Introduction}
\input{tex/1-intro}

\section{Related Work}
\input{tex/2-rw}

\section{Examining Gaze Depth Characteristics}
\label{section:3}
\input{tex/3-detection}

\subsection{Design Rationales Based on Gaze Depth Characteristics}
\input{tex/4-reqs}

\section{FocusFlow: Layer-based UI Design}
\label{sec:layer-based}

\input{tex/5-design}

\section{User Study}
\label{sec:userstudy}
\input{tex/6-userstudy}

\section{Case Study}
\label{sec:userstudy}
\input{tex/7-casestudy}

\section{Limitation and Future Work}

\input{tex/8-discussion}
\section{Conclusion}
\input{tex/9-conclusion}


\begin{acks}
This research was partially funded by UIUC-Insper program. We thank Professor Andrew Kurauchi and Professor Luciano Soares for their guidance in eye-tracking characterization and Arnav Shah for designing and implementing the surgical training application.
\end{acks}

\bibliographystyle{ACM-Reference-Format}
\bibliography{sample-base}


\appendix
\include{tex/10-appendix}

\end{document}

%% file: tex/1-intro.tex



Gaze interaction has gained popularity as an input method for 3D interaction in mixed reality (XR) headsets \cite{sidenmark2021radi, lu2021exploration, adhanom2023eye}, where the eye movements and gestures would be used as an input. 
Current gaze-based XR interaction methods often rely on gaze as a means of pointing, supplemented by other inputs such as hand gestures \cite{yu2021gaze}, head movement \cite{sidenmark2021radi}, or dwell time \cite{penkar2012designing, piumsomboon2017exploring} for selection confirmation to avoid the so-called "Midas Touch" problem \cite{velichkovsky1997towards} -- the problem of distinguishing intentional gaze inputs from involuntary fixations as the users perceive the scene. 
However, these works primarily utilize the direction of gaze and overlook the valuable \emph{visual depth} information, representing an additional free input dimension along the z-axis (Figure \ref{fig:teaser}). 
To see how, imagine a \textbf{window} through which you can see scenery outside. By bringing your gaze focus closer to the window, you blur out the distant scenery and are able to perceive dust particles on the window surface instead. Another example is autostereograms \cite{tyler1990autostereogram}, where slightly adjusting your visual depth allows you to perceive a 3D image (see example in Appendix \ref{Autostereogram}). \revision{In both of these examples, users actively manipulate their visual depth to select which object to observe. In this research, we aim to leverage this natural eye behavior and convert it into an intuitive and learnable interaction input in Virtual Reality (VR).}

\revision{More recent works on gaze-based VR/AR interaction demonstrate the great potential of visual depth as an interaction input to solve the Midas touch problem. These methods either guide the user to look at physical or virtual objects at different depths \cite{ahn2020verge,pai2016transparent,wang2022control, wang2022gaze, sidenmark2023vergence} or rely on voluntary eye convergence and divergence \cite{kirst2016verge, kudo2013input} by asking the users focusing on the nose or imagining to fixate on some point behind the display plane. \camready{These works prove the feasibility of leveraging visual depth as an input for VR headset, however, the usability and learnability of this new interaction design are not fully explored. Failure-prone and difficult-to-learn experience can lead to user frustration and fatigue, making the interaction unusable. How to guide the users to learn to manipulate their visual depth as a reliable input is still a research gap.}
}



\revision{
In particular, leveraging visual depth as an interaction input is challenging due to several practical factors. 
First, the natural eye movement is inherently unstable, resulting in noisy visual depth estimates. Moreover, the eye tracking modules in the headsets suffer from limited frame rate and jitters, adding extra noise to depth estimates. So, a usable interaction mechanism is required that is robust against imprecise eye input signal.
Second, user's visual depth fixation naturally relies on the presence of objects. So, it may feel counter-intuitive if the user is asked to transition to a depth without any visual referent. So, it is crucial to ensure that novice users can intuitively and effortlessly learn and use this interaction and a carefully designed learning procedure is needed to guide the users on how to master this new interaction mechanism. To tackle these challenges, we employed an empirical research approach. Initially, we devised a visual depth detection method for off-the-shelf VR headsets and conducted experiments to investigate the properties of visual depth input. Building upon the insights gained from these experiments, we subsequently developed our proposed user interface and incorporated a learning procedure with visual cues to help users adapt to the depth-control process.
}


\camready{As such, we present \emph{FocusFlow}, a novel gaze-depth interaction method that realizes the concept of \textbf{Virtual Window} (just like the examples mentioned above) by utilizing the 3D nature of immersive scenes and the innate human capability to shift focus across different visual planes. FocusFlow employs the eye trackers in VR headsets to estimate the user's visual depth. It then proposes a \emph{layer-based user interface}, where each layer acts as a transparent Virtual Window positioned at varying depths. These windows become interactive when a user's visual depth aligns with their location, allowing for seamless interaction without visual clutter. Specifically, FocusFlow selects interaction targets based on the user's gaze direction and activates Virtual Windows—initially hidden within the VR scene—through voluntary changes in visual depth (as illustrated in Figure \ref{fig:teaser}). This mechanism facilitates a range of applications, from quick previews and system panel activation to detailed zoom lenses for accessibility, enhancing the VR experience without overwhelming the user interface.
}


\revision{To help users master this new interaction mechanism, we incorporate \emph{visual cues} in the UI design and a learning process that allows users to establish a clear connection between the eye behavior and visual depth control, facilitating an intuitive and effortless interaction for novice users. The key intuition is that users can develop muscle memory for freely shifting their focus between different visual planes even if empty of objects. We propose two learning strategies, \emph{in-stages learning}, and \emph{adaptive learning}. The in-stages learning uses a series of \emph{strong} to \emph{weak} visual cues at a certain depth (e.g. where the transparent virtual window is located) so the user can practice visual depth control by just moving her gaze from the target object to the cue. While these cues are effective in practicing visual depth control, they also cause visual interference for the user during non-interaction times. The proposed \emph{adaptive} visual cues addresses this issue by adaptively changing the transparency of visual cues as the user gets closer to the intended visual depth. The transparency level of the cue also serves as a feedback for the user to better control their visual depth changes. To help master this interaction, we design the adaptive learning process gradually reduces the guided depth range until reaching no visual cue.}

\camready{
To evaluate the usability and learnability of FocusFlow, we conducted quantitative and qualitative evaluations, considering input efficiency, accuracy, and cognitive load. Results from 24 participants indicate that FocusFlow offers an efficient, easy-to-learn experience for hands-free selection tasks, effectively addressing the Midas Touch problem while maintaining natural interaction. Furthermore, our findings suggest that the adaptive visual cues in our design enhance users' depth perception during visual depth transitions. Additionally, the adaptive learning strategy improves users' proficiency with our gaze-depth interaction method.}

\camready{The contribution of this work comprise:
\begin{enumerate}
\item We analyze the characteristics of gaze-depth estimates and propose strategies for utilizing this imprecise input information for interaction design.
\item We develop a layer-based user interface for gaze-depth interaction by realizing the concept of the ``Virtual Window''. This interface supports intuitive and robust hands-free selection in VR while avoiding Midas Touch problem.
\item We design three visual cues that provide users with depth perception, and two learning strategies, in-stages learning and adaptive learning, to guide novice users in adapting to depth control and developing muscle memory for gaze-depth interaction.
\item We conduct a user study and a case study to assess both usability and learnability of our proposed system, and compare the effectiveness for different learning strategies.
\end{enumerate}
}

%% file: tex/2-rw.tex
In this section, we first summarize different eye behaviors and the corresponding input techniques which can be used for gaze interactions. 
Based on this, we then review the evolution of gaze-based interaction in VR and some representative methods to identify the research gap.
Lastly, we discussed design space research on user interfaces that integrate gaze input.

\subsection{Eye Behaviors and Input Techniques}
\label{Eye Behaviors and Input Techniques}

\paragraph{\textbf{Eyelid Functions}}
Eyelid movements, ubiquitous in daily life, also hold value as input mechanisms. \textbf{Blinking} and \textbf{winking}, rapid opening and closing of the eyelid, serve to distribute tears across the eye's surface. Despite their inherent simplicity, these actions can be random and noisy, complicating the balance between efficiency and accuracy. Consequently, most input techniques leveraging blinking and winking are geared towards accessibility applications~\cite{krolak2012eye}. Another area of interest is the detection of \textbf{eyelid openness} \cite{arvacheh2006iris}, a measurable value that can be used to control objects \cite{yi2022deep}.

\paragraph{\textbf{Eye Movements}}
Eye movements have the potential to serve as input signals, as they can reflect attention direction. Notably, \textbf{fixation} is the act of maintaining a steady gaze on a specific location, while \textbf{saccade} involves quick, darting eye movements that suddenly shift the fixation point. Feit et al. \cite{feit2017toward} explored the utilization of saccades and fixations for accurate interaction input, laying the groundwork for many gaze-based interactions. \textbf{Smooth pursuit movements} allow our eyes to steadily track a moving object. Inspired by this, researchers have proposed several pursuit-based gaze input methods for tasks such as selection \cite{esteves2015orbits} and text entry \cite{zeng2018text, abdrabou2019calibration}.

\revision{\paragraph{\textbf{Vergence}}
Vergence is a specific type of eye movement that represents the coordinated movement of both eyes in opposite directions to sustain binocular vision As a result of such eye movement, the user's visual depth will change. Previous works such as Verge-it \cite{ahn2020verge}, employ vergence eye movement as an interaction input to select objects. In these techniques, the user is either asked to follow the motion of an object which results in vergence changes, or to look at their nose to create voluntary vergence control. We build our proposed interaction method upon these eye movements and vergence control systems, but we propose an intuitive user interface design that allows the user to perceive \textbf{visual depth} and actively use it for interaction with minimal cognitive load or visual interferences. }




\subsection{Gaze-Based VR Interaction}
\label{Gaze-based Interaction}

\vspace{.3em}\noindent\revision{\textbf{{Gaze-ray-based Interaction.}} Eye movements were first utilized for VR pointing tasks by Tanriverdi and Jacob \cite{tanriverdi2000interacting}, resulting in faster interaction compared to hand-based gesture methods. Building on this, Sidorakis et al. \cite{sidorakis2015binocular} introduced a VR keyboard input method that employs gaze pointing, leading to fewer typing errors than conventional interfaces. Mardanbegi et al. extended this approach with EyeSeeThrough \cite{mardanbegi2019eyeseethrough}, a VR selection method that aligns the intended object with a tool menu using the user's gaze ray. Piumsomboon et al. \cite{piumsomboon2017exploring} further explored the application of gaze direction in VR, proposing different \emph{gaze ray selection} methods that only leverage the direction of gaze.
However, a direction-only gaze selection method causes a so-called Midas Touch problem \cite{jacob1990you}, where unintentional inputs might be recognized as commands. }

\vspace{.3em}\noindent\revision{\textbf{{Dwell-time-based Gaze Interaction.}} Researchers have attempted to mitigate Midas Touch problem by incorporating dwell time \cite{lu2021itext, lu2021evaluating, majaranta2009fast, mott2017improving, mardanbegi2019eyeseethrough,pfeuffer2020empirical} or similar attention patterns \cite{piumsomboon2017exploring}. While these \textbf{dwell-based gaze selection} techniques can reduce false triggers, they also slow down the input response due to the added time required for dwell validation, which degrades the interaction efficiency and user experience.} 

\vspace{.3em}\noindent\revision{\textbf{{Multi-modal Interaction.}} Another approach for addressing the Midas Touch problem is to expand the input dimension 
such as leveraging \emph{multi-modal inputs}, which combines gaze with another input modality. The typical multi-modal inputs combine the gaze direction and controller inputs \cite{sendhilnathan2022detecting, mardanbegi2019eyeseethrough}. Other works combine gaze and head information. For example, Sidenmark et al. proposed Radi-Eye \cite{sidenmark2021radi}, which combines head and eye movement to validate selections. Wei et al. \cite{wei2023predicting} proposed a prediction method for target selection in AR based on the distribution of eye and head endpoints. More recent papers explore the combination of gaze and pinch \cite{pfeuffer2017gaze}, gaze and hand alignment\cite{lystbaek2022gaze}, humming \cite{hedeshy2021hummer}, or other hand gestures \cite{yu2021gaze, lu2021exploration, lu2021gaze, hou2023classifying,bao2023exploring}. In these systems, the pointing information is extracted from eye movement and selection or confirmation of information are done with hand gesture. The gaze and hand multi-modal interactions are also built into commercial products such as Apple Vision Pro. However, involving other modalities inevitably increases the interaction complexity and may not be applicable in certain applications that require hands-free interaction.} 

\revision{An alternative way to expand interaction dimension is to leverage \textbf{more eye-input information}. For instance, unconventional blinking \cite{lu2021exploration,rebsdorf2023blink,yu2021gaze} is used as a complementary gaze input to direction, but the proposed technique suffers from unintentional triggering. Yi et al. introduced DEEP \cite{yi2022deep} which leverages eyelid movement as a new input for controlling to allow users to adjust the visual depth in a scene, thus improving the gaze-pointing accuracy when some objects are occluded. However, this approach relies on dwell time, which still suffers from efficiency and usability problems. In this study, we incorporate visual depth as a novel input dimension in gaze interaction to address the Midas Touch problem in gaze selection, while maintaining interaction efficiency.}

\vspace{.3em}\noindent\revision{\textbf{{Gaze-depth-based Interaction.}}
Gaze vergence is an emerging eye-input information that has been used for VR interaction. Compared to multi-modal gaze interaction methods that leverage handheld controllers or hand gestures, visual depth offers a completely hands-free interaction mechanism. Previous works have shown the feasibility of visual depth estimations in the VR headsets by using binocular disparity and stereoscopic vergence in VR headset \cite{hirzle2019design, vidal2014looking}. The concept of visual depth as a new interaction input has explored in the previous works, either by defining a semi-transparent window at a different focal depth \cite{pai2016transparent, wang2022control, wang2022gaze}, tracking voluntary vergence movements \cite{kirst2016verge, ahn2020verge}, or matching the vergence changes with the depth changes of a moving objects \cite{sidenmark2023vergence} in VR. 
\camready{
These studies have confirmed the efficacy of the gaze-depth-based interaction method from a technical standpoint. However, the learnability of this interaction poses a significant concern, as most users are not acquainted with gaze-depth input. There is a clear need for specific designs that instruct users on how to actively adjust their visual depth, along with comprehensive user studies to assess the usability and efficiency of this novel interaction mechanism.
} 
This paper targets these limitations and proposes a layer-based UI design to leverage active visual depth manipulation despite unstable depth estimates of existing eye tracking systems. In addition, we highlight an integrated learning process that leverages visual cues to help users learn how to voluntarily shift their focal depth. We also perform a user study and a case study to evaluate user experience and mental model in our gaze-depth interaction process.
}



\revision{\subsection{Immersive Gaze-based UI Design}
There is a flurry of work on user interface design that incorporate gaze input in VR and AR systems. For example, Pfeuffer et al. \cite{pfeuffer2021artention, blattgerste2018advantages} proposes a design space for gaze-input based interfaces for AR applications. Hirzle et al. \cite{hirzle2019design} introduced a design space for head mounted devices focusing on human depth perception and technical issues. In this paper, we propose a layer-based UI design for virtual reality that integrates active visual depth manipulation and gaze movement for VR interaction. We evaluate this new interaction and demonstrate the potential of this gaze-only interaction method as an effortless and intuitive method for everyday VR through a learning process. }

%% file: tex/3-detection.tex
In order to design an intuitive and robust gaze-depth interaction, we first conducted a user study to explore the performance of built-in eye trackers in commodity VR headsets and visual depth characteristics. 

\begin{figure}
 \centering
 \begin{subfigure}[b]{0.4\textwidth}
     \centering
     \includegraphics[width=\textwidth]{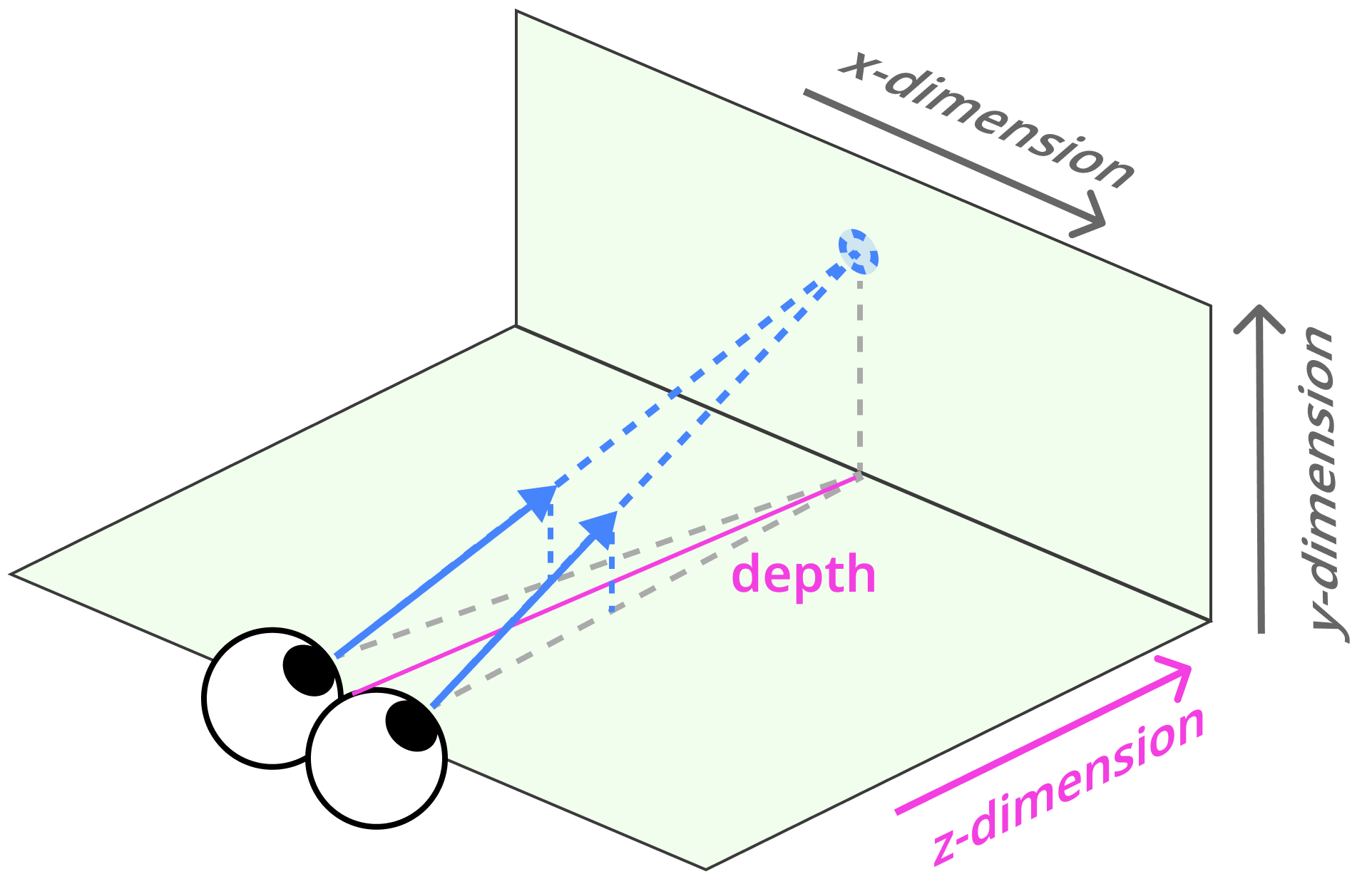}
     \caption{Visual depth calculation. Two gaze rays are projected onto the x-z plane. The visual depth is calculated as the distance between the intersection of projected gaze rays and the midpoint of the user's eyes.}
     \label{fig:depthDetect}
 \end{subfigure}
 \hfill
 \vspace{.1in}
 \begin{subfigure}[b]{0.45\textwidth}
     \centering
     \includegraphics[width=\textwidth]{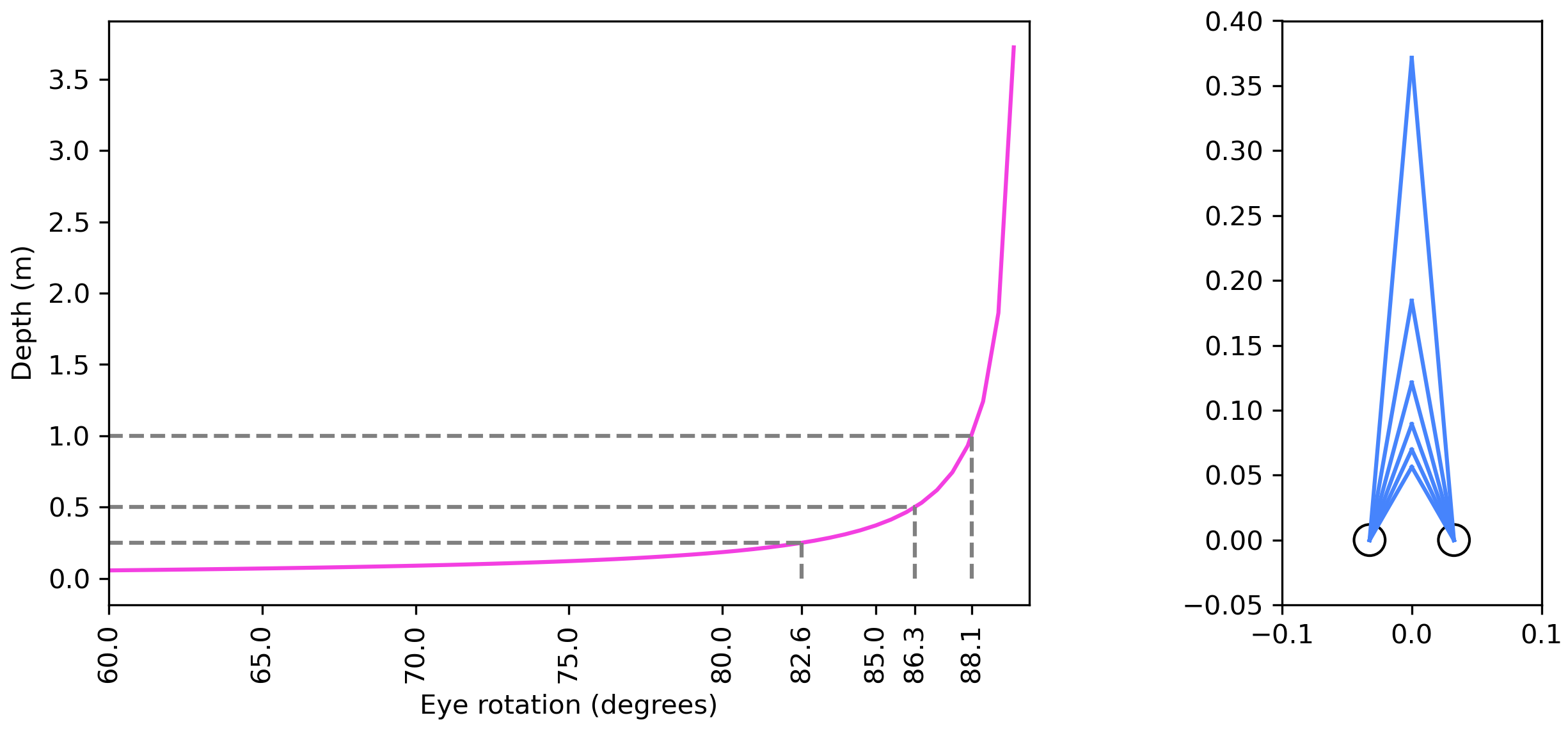}
     \caption{Gaze-depth Sensitivity analysis. For a fixed interpupillary distance, a small change in gaze direction at larger angles (aka larger depths) create much larger changes in visual depth. Therefore, larger visual depths are more prone to noises. }
     \label{fig:sensitivity analysis}
 \end{subfigure}
  \caption{Depth calculation and sensitivity analysis.}
\label{fig:dectionAll}
\end{figure}

\subsection{Visual Depth Estimation}

\revision{ In order to calculate the user's visual depth, we use the built-in eye trackers in VR headsets, which capture the positions of eyeballs and converts them into binocular gaze rays originating from eyes and intersecting at a specific depth (see Figure \ref{fig:depthDetect}). However, two rays in 3 dimensions may not intersect at a point specially with a small error in gaze direction estimations. To address this issue, we leverage the projection of these two rays onto a plane which ensures the intersection of the two rays. We define the projection plane along the two gaze ray origins (x-axes) and z-axes. As such, the distance between the eye line and the intersection point in this projection plane will be equal to the visual depth. To deal with random eye movements \cite{mughrabi2022my, zhang2021evaluating} and the resulting depth jitters, we apply a moving average technique with 0.2 seconds of time window to filter out the high frequency depth fluctuations in the depth estimates.} 

\begin{figure}[htb]
  \centering
  \includegraphics[width=\linewidth]{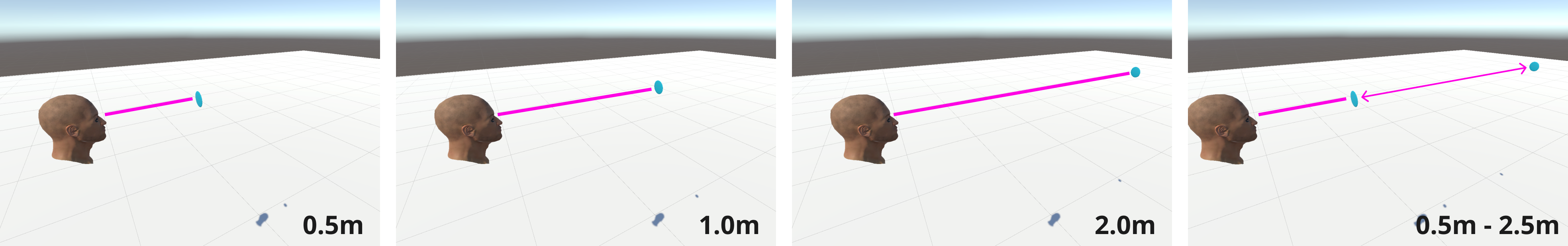}
  \caption{Experiment settings. The participants are required to look at three static objects located at 0.5m, 1.0m, and 2.0m away from themselves. Then they are asked to follow a moving target moving back and forth between 0.5m and 2.5m away with their gaze.}
  \label{fig:pre-experiemnt}
\end{figure}

\subsection{Experiment Design For Pilot Study}
\label{subsec:detection_settings}

For the apparatus, we use an HTC VIVE Pro Eye headset, which has an FOV of 100$^{\circ}$ and a 90Hz frame rate. The eye tracking data is recorded by a built-in eye tracker camera at a frame rate of 120Hz and 0.5-1.1$^{\circ}$ tracking precision. To evaluate the accuracy of visual depth estimates, we invited 7 participants from the campus. For each test, the participant is sitting on a chair while wearing the headset and follows the instructions on looking at certain targets.  

We designed two sub-tasks in this experiment. In the first task, participants are asked to stare at a static VR target for five seconds located straight ahead at different distances from the user (0.5m, 1.0m, and 2.0m respectively). The target will be highlighted to capture participants' attention. In the second task, they are asked to look at a moving target that moves back and forth toward the user at depths between 0.5m to 2.5m. The participants are allowed to blink and take a rest if they feel tired in both stages. Figure \ref{fig:pre-experiemnt} shows all the tasks in the experiment.

\begin{figure*}[t]
  \centering
  \includegraphics[width=\linewidth]{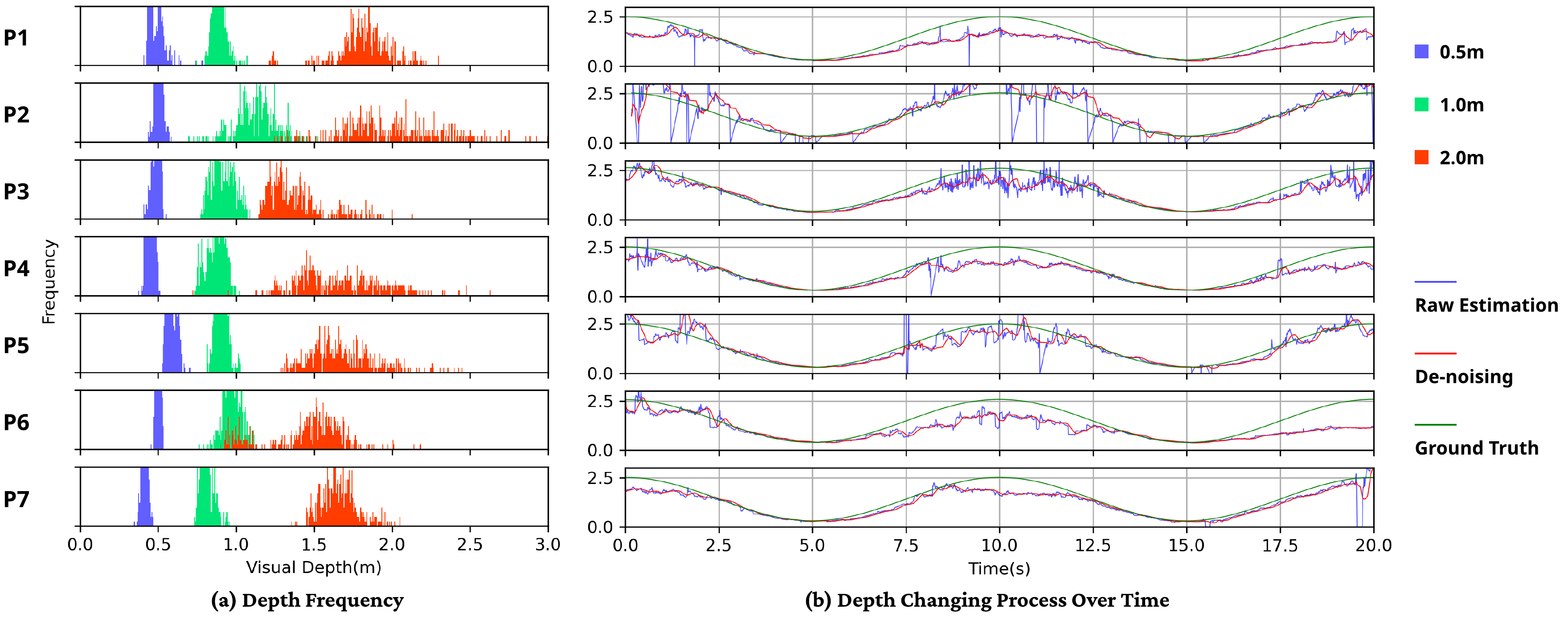}
  \caption{Visual depth detection results. (a) per-participant histogram of estimated depths for three static targets at different distances; (b) snapshot of estimated depth fluctuations for the moving target with and without de-noising.}
  \label{fig:evaluation_detection}
\end{figure*}

\subsection{Takeaways}
\label{detection_analysis}

Figure \ref{fig:evaluation_detection} summarizes the results of the visual depth estimates and visual depth behaviors for the 7 participants. Figure \ref{fig:evaluation_detection}a shows the histogram of estimated depths for task 1 with static targets at different depth, represented by different colors. Figure \ref{fig:evaluation_detection}b shows snapshots of depth estimates in task 2 (moving target) with and without denoising. We conclude the following visual depth characteristics:

\begin{enumerate}[start=1,label={C.\arabic*},wide = 0pt, leftmargin = 1.9em]

\item \textbf{Visual depth is a noisy but usable signal:} In Figure \ref{fig:evaluation_detection}a, we can see that estimated visual depths are always presented as a distribution around the target's groundtruth depth. This verifies the unavoidable random eye movements even when looking at a single static object. Nonetheless, we are able to find a relatively clear demarcation between the blue distribution (target at 0.5m distance) and the red distribution (target at 2m distance). This suggests that visual depth can be distinguished at relatively coarse granularity, but it is still a usable signal.
    
\item \textbf{Estimated Visual depth accuracy decreases with the increase of distance:} By comparing the visual depth patterns in Figure \ref{fig:evaluation_detection}a and b at different distances, we can conclude that visual depth can be estimated with high accuracy (absolute depth values) and precision (standard deviation of depth distributions) in short distances, but both the depth estimate variations and accuracy drops for larger depths. We then conduct a theoretical sensitivity analysis of the visual depth to the eye rotation and gaze direction. As shown in Figure \ref{fig:sensitivity analysis}, the visual depth becomes more sensitive to the eye rotation changes at larger eye rotation values. So, at longer distances (more tahn 2 meters away), even a small amount of noise or error in gaze direction estimates can result in much larger error in visual depth calculations. 


\item \textbf{De-noising method plays a critical role.} The raw depth estimation curve (red) in Figure \ref{fig:evaluation_detection}b shows persistent fluctuations, especially when the target distance is larger than one meter from the user. After applying our de-noising method, the new depth curve (red) is smoother and drastic depth fluctuations due to random eye movements are eliminated. This indicates that our de-noising method plays a crucial role in pre-processing the input visual depth data.

\end{enumerate}


%% file: tex/4-reqs.tex


\revision{
Our pilot study for characterizing the visual depth estimation sets certain trade-offs on leveraging gaze depth as an interaction input. In one hand, gaze depth estimates are mainly reliable in short distances, with larger error in estimated depths as the distance increases. So, for long distance operations, we can only monitor obvious depth shifts which limites the gaze-depth inputs to coarse grained depth changes only. 
On the other hand, gaze-based interaction is more useful in large virtual environments for interacting with distant objects, where other modalities (e.g. hand or head gestures) cause user fatigue  \cite{yu2021gaze}. So, the main objective of this work is to break this trade-off by taking into account users' spatial perception, input efficiency, and user's learning ability.} As such, we define three design goals for FocusFlow. Using FocusFlow, :


\begin{enumerate}[start=1,label={D.\arabic*},wide = 0pt, leftmargin = 1.9em]
\item \textbf{Users should be able to achieve hands-free selection efficiently and accurately, without encountering the Midas Touch problem.} The efficiency of a gaze-based interaction is measured by the wait time between the user input and the system response and is expected to be below the user's perceptible limit. The accuracy is indicated by the success rate and false trigger rate of an operation (e.g. selection), which means the selection and deselection intention should be accurately captured by the system.


\item \revision{\textbf{Users should be able to perceive the depth and accordingly adjust their focal depth for interaction.} Despite the difficulty of users in adjusting their visual depth to a certain length without any reference in space, perceiving the depth by using alternative visual cues can improve voluntary gaze depth changes. The system should offer an intuitive mechanism to grasp the concept of visual depth transitions.} 

\item \revision{\textbf{Users should be able to develop muscle memory to provide efficient and effortless interaction.} An immersive virtual scene should be compatible with user instincts to avoid user discomfort \cite{weech2019presence}. Since voluntary gaze-depth change is a brand-new input method for general VR users, learnability is critical to our interaction approach. Some guidance and feedback should be provided to facilitate the user's learning process.}
\end{enumerate}

\section{FocusFlow: System Overview}
This paper proposes FocusFlow, which consists of three main components to satisfy the design goals mentioned in the previous section:

\begin{itemize}
    \item \revision{\textbf{Layer-based UI Design:} FocusFlow presents a layer-based UI design for virtual reality applications that supports gaze-depth interaction. In this design, interactive virtual contents are organized into distinct spatial layers at different visual depths relative to the user. The visibility indexes of these layers are defined based on the user's gaze depth, transitioning between completely opaque when the user's gaze depth matches with the depth of a layer to completely transparent otherwise. Section \ref{sec:layer-based} explains how FocusFlow avoids Midas Touch problem by carefully selecting the layers in the UI and active gaze-depth manipulation to activate a layer. }
    \item \revision{\textbf{Gaze-depth Interaction Design:} FocusFlow leverages the proposed layer-based UI design and offers a gaze-only interaction mechanism by combining gaze direction and gaze depth. The gaze direction is used for selection of the target object for interaction and gaze depth is used for activating a certain layer in the UI. Section \ref{sec:design_applications} discusses a variety of applications that can benefit from this hands-free gaze-only interaction design.}
    \item \revision{\textbf{Adaptive Interaction Learning Process:} FocusFlow takes advantages of a series of visual cues to help users master this brand-new interaction. The visual cues are designed such that they provide users with depth perception and learning opportunities to practice, receive feedback, and master intuitive adjustment of their gaze depth for interaction. Section \ref{sec: Learning Strategy} describes two different learning strategies based on in-stages and adaptive visual cues.  }
\end{itemize}

%% file: tex/5-design.tex

\begin{figure}[htb]
  \centering
  \includegraphics[width=1.0\linewidth]{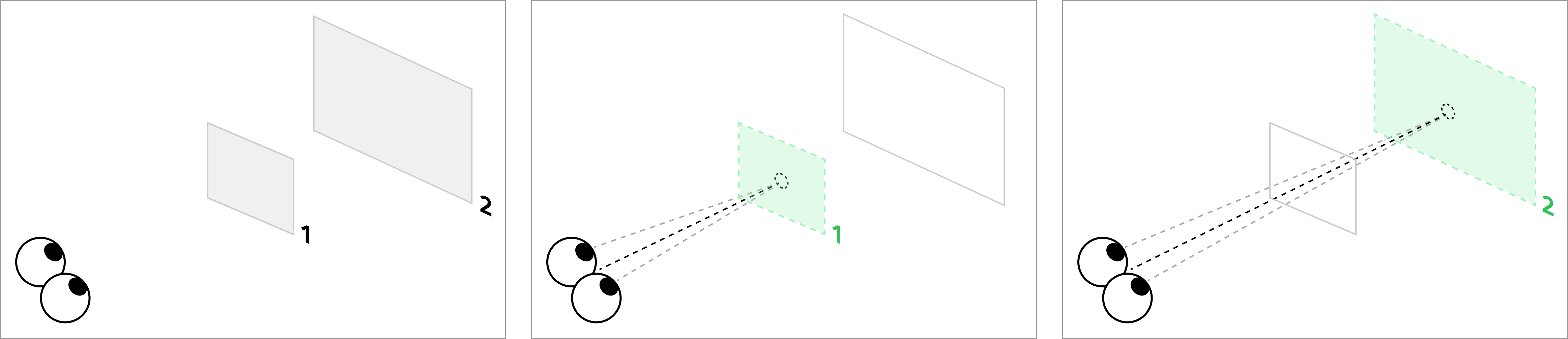}
  \caption{Layer-based UI. The layers are arranged at different depths along the z-direction. The user can match the corresponding layers by changing the visual depth. The matched layers will be activated.}
  \label{fig:layers}
\end{figure}


Our vision perception undergoes a significant transformation when there is a change in visual depth. More specifically, a change in the perceived depth causes objects within the original field of view to appear out-of-focus and blurred since they are no longer aligned with our current focal plane. Our goal is to mimic the same intrinsic user experience in the virtual environments by aligning the in-focus objects with the user's perceived depth. 
To achieve this, we propose a ``layer-based'' spatial user interface that organizes virtual content into distinct layers. We also propose an activation logic that defines the visibility index of each layer based on the user's visual depth. These visibility indexes are dynamically adjusted based on shifts in the user's visual depth. To create an immersive experience, the layer that is matched with user's visual depth is visible with low transparency so that the user can perceive its position relative to further layers, while the closer layers are completely invisible. As such, we can enhance the user's sense of depth perception. Figure \ref{fig:layers} shows a high-level overview of the proposed layer-based UI.

\begin{figure*}[htb]
  \centering
  \includegraphics[width=0.7\linewidth]{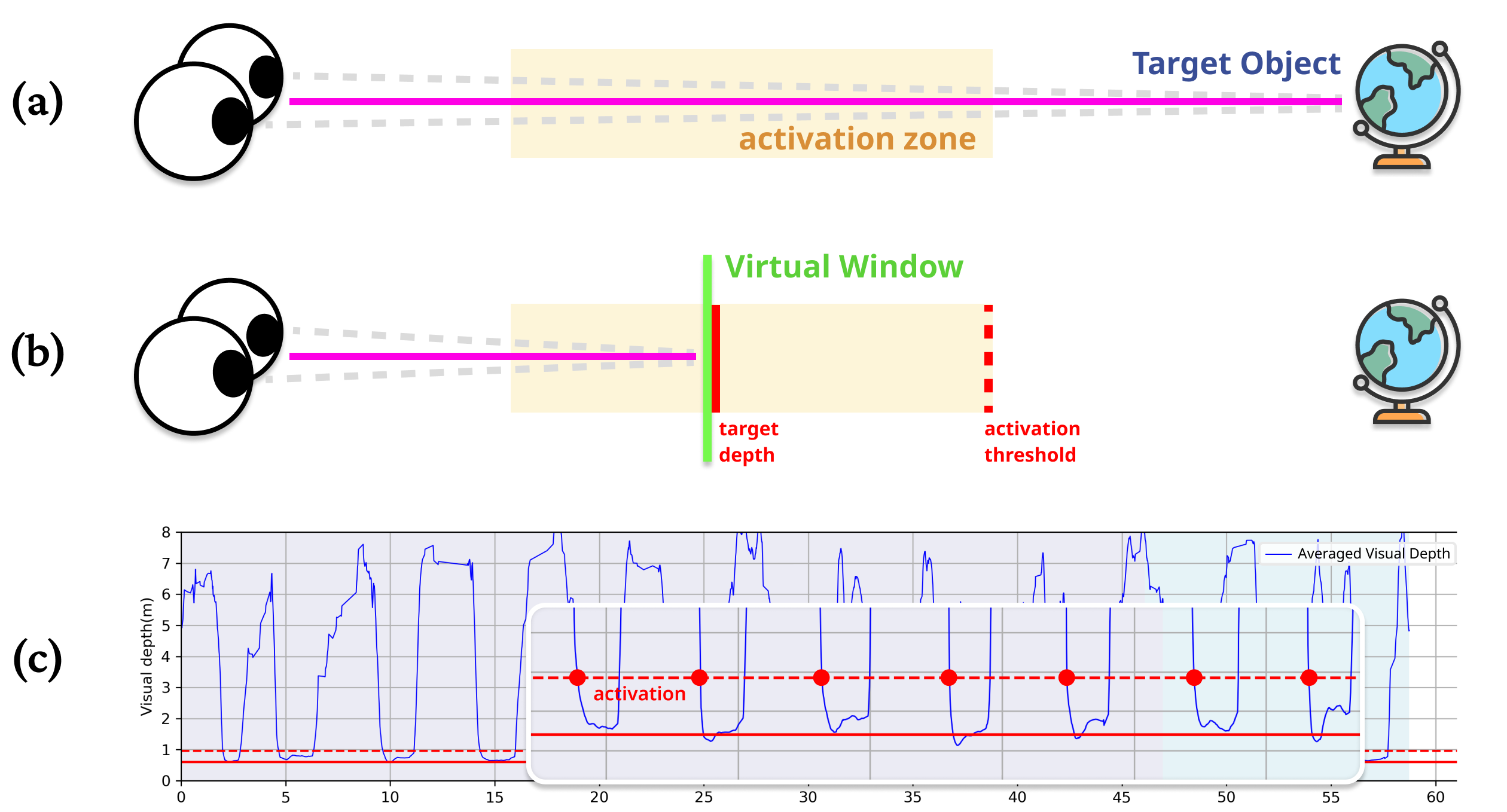}
  \caption{Activation logic. (a) Pointing: When the user is looking at the target object in VR environment, the gaze point falls on the object and the visual depth exceeds the activation zone. (b) Depth Shift: When the user's visual depth falls in the activation zone, the Virtual Window will be activated. (c) The gaze depth change in multiple activations.}
  \label{fig:activation_logic}
\end{figure*}

\subsection{Virtual Window: A Depth-based Selection Widget}



Leveraging the proposed layer-based UI, we design \textit{Virtual Window}, a hands-free selection widget with visual depth input. The high level idea is to replicate the user experience of seeing through a ``window'' or looking at the ``window'' itself in the virtual environment. Our vision system is capable of ignoring the presence of window and only focus on scenery behind it. At the same time, by bringing your visual focus closer to the window, you blur out the distant scenery and zoom into dust particles on the window surface instead. 

Similarly, this virtual Window is an information layer, floating in front of the user, and it is invisible by default. The user can activate the window by bringing their gaze point closer. When the visual depth is matched with the window distance, the information on the Virtual Window will be activated, as shown in Figure \ref{fig:activation_logic}. This transition process is highly intuitive, as it provides users a sense of ``grabbing in'' or ``taking a closer look'' at detailed information when they bring the visual depth closer. Similarly, to send the window to its default transparent status, the user only need to push their visual depth further, back to the portal layer. The window's visibility change is a response indicating the activation is successful, and also naturally helps users to maintain the visual depth as they are looking at information on the window.

\revision{One challenge with this layer-based UI design is potential false triggers or no triggers due to gaze-depth estimation errors. To avoid false triggers, we employ our findings in the pilot study, described in Section 3, that showed a higher depth estimation accuracy at shorter distances. Therefore, we select the virtual window depth within the 2m range of the user at a distance that no virtual object is present. To avoid false negatives and no triggers, we define an \emph{activation zone}, shown in Figure \ref{fig:activation_logic}, which is a predefined spatial range around the actual virtual window depth. If the user's visual depth falls inside this activation zone, the virtual window will be activated. }




\begin{figure}[htb]
  \centering
  \includegraphics[width=1.0\linewidth]{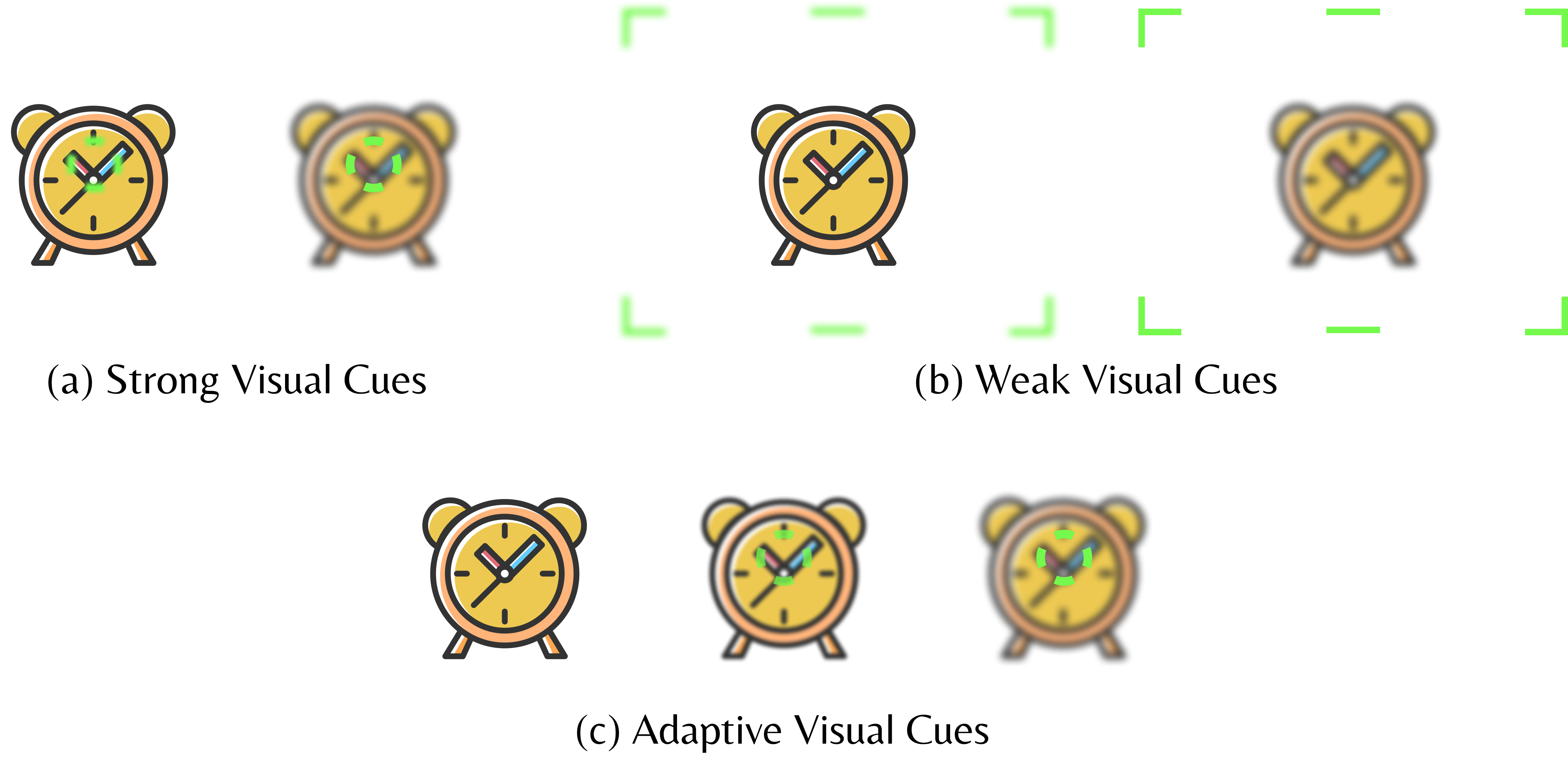}
  \caption{Three different visual cues. (a) Strong visual cues appear at the \textbf{center} of the view with \textbf{constant} low transparency. (b) Weak visual cues appear at the \textbf{margin} of the view with \textbf{constant} low transparency. (c) Adaptive visual cues will appear at the \textbf{center} of the view with \textbf{variable} transparency which depends on the user's gaze depth.}
  \label{fig:visual cues}
\end{figure}

\subsection{\revision{Visual Cues}}
While visual depth transition comes naturally to our vision system, it highly relies on the presence of physical/virtual content at that particular depth. So, it may be at first counter-intuitive for the users to perceive the depth and switch their gaze to a point in the space with no visual feature. \revision{Previous works on gaze-depth interaction address this issue by asking users to look at their nose to create vergence \cite{kirst2016verge, ahn2020verge}. However, this may results in user discomfort and fatigue. Instead, FocusFlow defines three types of \textbf{visual cues} that offer intuitive interaction with minimal visual interference.}

\revision{The first design is a \textit{strong guidance} (Figure \ref{fig:visual cues}a), which is a static green circle in the center of the user's view at the close depth. With this, users can transition their visual depth from far to nearby by directly looking at the circle. However, since the circle is always in the user's field of view, it may be a strong visual interference for some users. The second design is a frame layout in the corners of the view, called \textit{weak guidance} (Figure \ref{fig:visual cues}b). Users can perceive the layer depth through their peripheral vision to transition their visual depth. As such, the strong guidance is a direct referent for the gaze point to focus, but the weak guidance is an indirect referent to the layer depth. The third design is an \textit{adaptive guidance} (Figure \ref{fig:visual cues}c), which is also a circle in the center, but its visibility actively changes based on the user's visual depth. We map its opacity to a certain range of the user's visual depth. When users are looking at distant objects, the circle is not visible so there is no visual interference. As users look closer, its opacity increases so users can find the circle and manipulate their gaze depth easier. In this adaptive process, the visual cue is not only a depth referent but also an indicator of the user's visual depth, which could serve as a feedback to the user; from the opacity or transparency of the circle: the closer the visual depth, the more obvious the circle will be. This helps users understand the gaze-depth interaction mechanism and better perceive the depth information in virtual scenes. }

\begin{figure}[htb]
  \centering
  \includegraphics[width=1.0\linewidth]{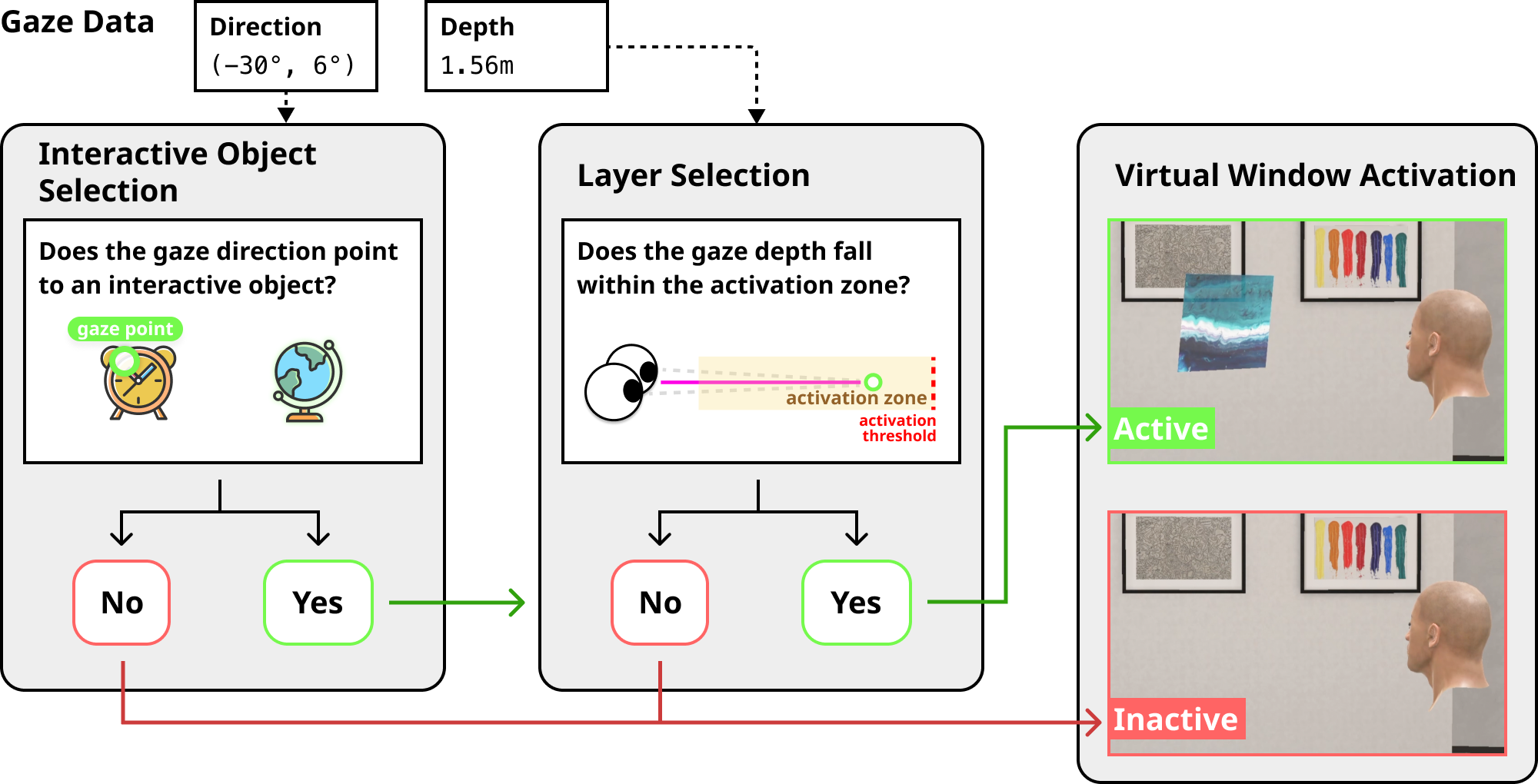}
  \caption{System pipeline of FocusFlow. Eye Tracker provides gaze data to the Interactive Object Selection and Layer Selection modules, and these two modules then send signals to control the activation of the Virtual Window.}
  \label{fig:pipeline}
\end{figure}

\section{\revision{FocusFlow: Interaction Design}}

\revision{Figure \ref{fig:pipeline} illustrates the overall control logic of FocusFlow's interaction design. The process is initiated with the eye tracker module in the headsets, which continuously monitors the user's binocular gaze directions. From the gaze direction information, it can then compute the gaze depth to determine the exact point the user is looking at (as described in Section 3.1). The Eye Tracker also employs a 'moving average' de-noising technique to refine the gaze data, ensuring that the data is smooth and free from random fluctuations that could lead to depth estimation errors.}

\revision{The gaze data is then passed to the \emph{Interactive Object Selection} module, which determines whether the user's gaze is aligned with an interactive object in the environment. The module mainly searches for intersection of the gaze rays with individual objects in the virtual scene and moves forward to layer selection module if the an object is selected.}

\revision{Upon selection of the interactive object, FocusFlow continuesly tracks the gaze depth estimates and evaluates whether the visual depth coincides with the activation zone, which is a predefined spatial area where user interaction is intended to occur. If the visual depth falls outside this zone, it is interpreted as a non-interactive state. However, if the gaze depth is within the zone, indicated by a "Yes" decision, it triggers the interaction signal.}

\revision{Upon satisfying both the direction and depth conditions, the interaction signal is then conveyed to the Virtual Window Activation module, which is responsible for managing the active window elements on the user interface. It reacts to the interaction signal by activating or deactivating window elements. For instance, if the user's gaze is directed at and focused on an interactive object within the activation zone, the corresponding virtual window element becomes active with full visibility. Conversely, if the user's gaze shifts away, the virtual window will disappear and return to an 'Inactive' state, as depicted in Figure \ref{fig:pipeline}.}

\begin{figure}
 \centering
 \begin{subfigure}[b]{0.23\textwidth}
     \centering
     \includegraphics[width=\textwidth]{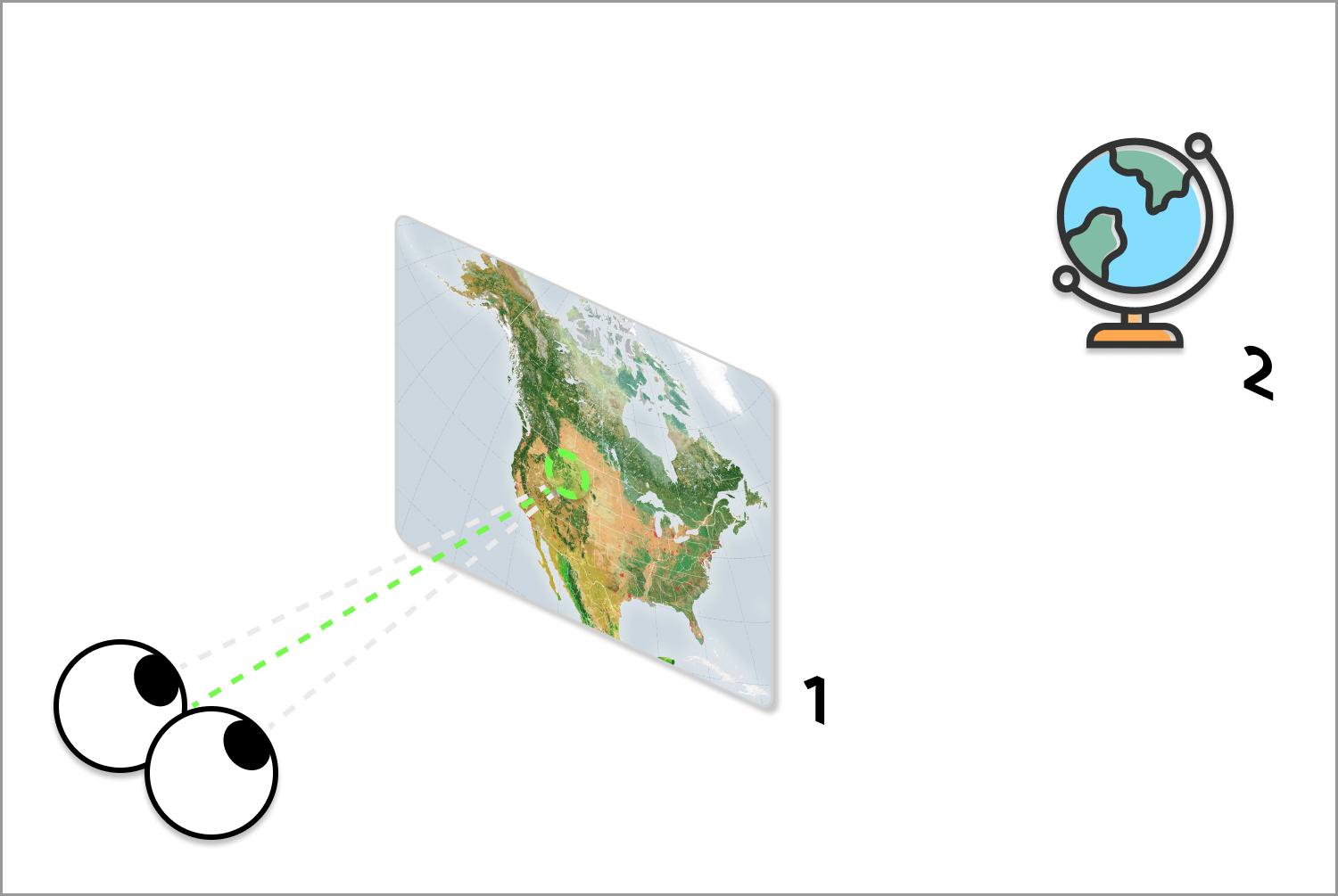}
     \caption{Quick Preview}
     \label{fig:app-a}
 \end{subfigure}
 \hfill
 \begin{subfigure}[b]{0.23\textwidth}
     \centering
     \includegraphics[width=\textwidth]{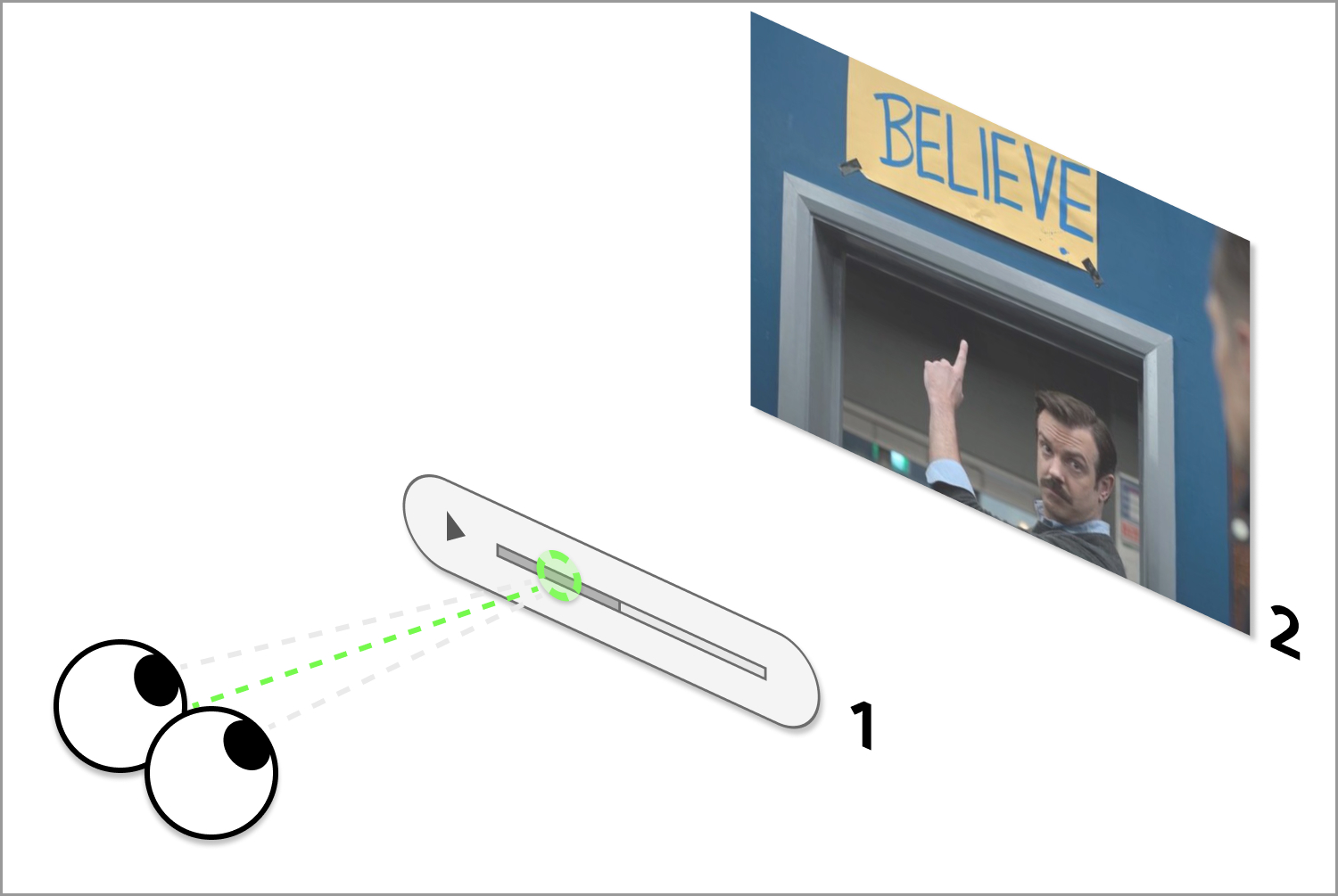}
     \caption{Safe Activation}
     \label{fig:app-b}
 \end{subfigure}
 \hfill
 \begin{subfigure}[b]{0.23\textwidth}
     \centering
     \includegraphics[width=\textwidth]{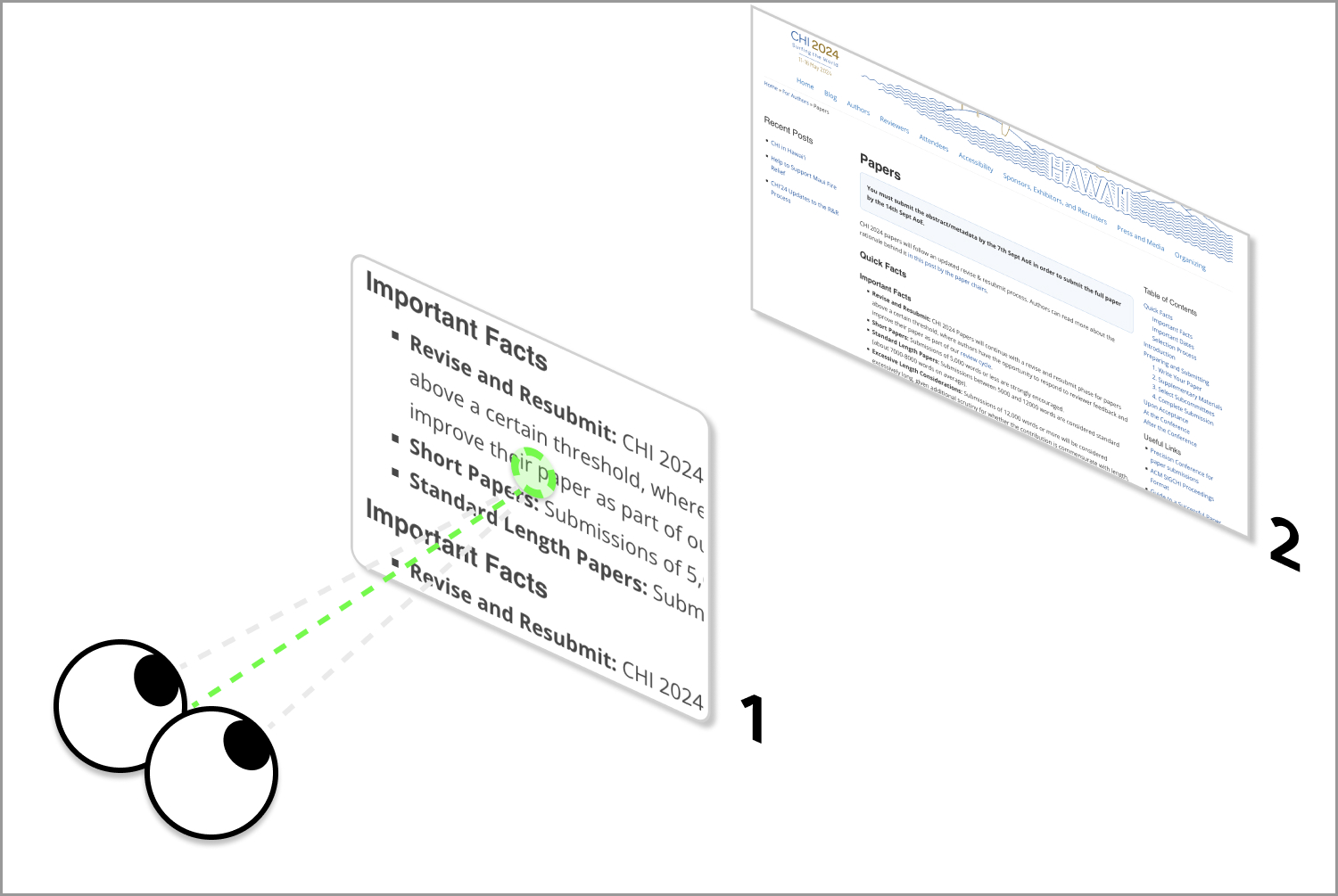}
     \caption{Zoom Lens}
     \label{fig:app-c}
 \end{subfigure}
\caption{Example applications of Virtual Window.}
\label{fig:application}
\end{figure}

\subsection{Interaction Applications}
\label{sec:design_applications}

\revision{The proposed layer-based UI and interaction offers a flexible design that can support different functionalities for the Virtual Windows. We implement and demonstrate FocusFlow functionality across several example applications:} 

\vspace{.3em}\noindent\textbf{Quick Preview.} \revision{Given the expected short depth transition delays in our visual systems (i.e.160-200 milliseconds \cite{10.1093/med/9780199969289.003.0009}),} users can activate the Virtual Window very fast, making it suitable for quick information preview. As Figure \ref{fig:app-a} shows, we can display preview information about the selected item on the virtual window, which is very common in scenarios like gaming, training, 3D creation, virtual tours, etc.

\vspace{.3em}\noindent\textbf{Safe Activation.}
Compared to actions such as gaze direction change, blinking, and finger movement, wide range visual depth shift is a relatively low-frequency and low-randomness action. Therefore, we can apply the depth activation method to scenarios where we don't want false touches to occur, which we call ``safe activation''. For example, when browsing a web page or watching a video in full screen, we can activate a toolbar by shifting the visual depth. Since our visual depth is relatively stable, we can avoid false touches that could interfere with our viewing experience. (Figure \ref{fig:app-b})

\vspace{.3em}\noindent\textbf{Zoom Lens.}
Visual depth is an input dimension along z-axis, where occlusion happens. This constraint actually offers logic connection between different layers, for example, the Virtual Window can also be used as zoom lens to magnify the object behind it. (Figure \ref{fig:app-c}) This design connects the information layer (Virtual Window) and portal layer (environment), i.e. providing a zoomed view.

\begin{figure}[htb]
  \centering
  \includegraphics[width=1\linewidth]{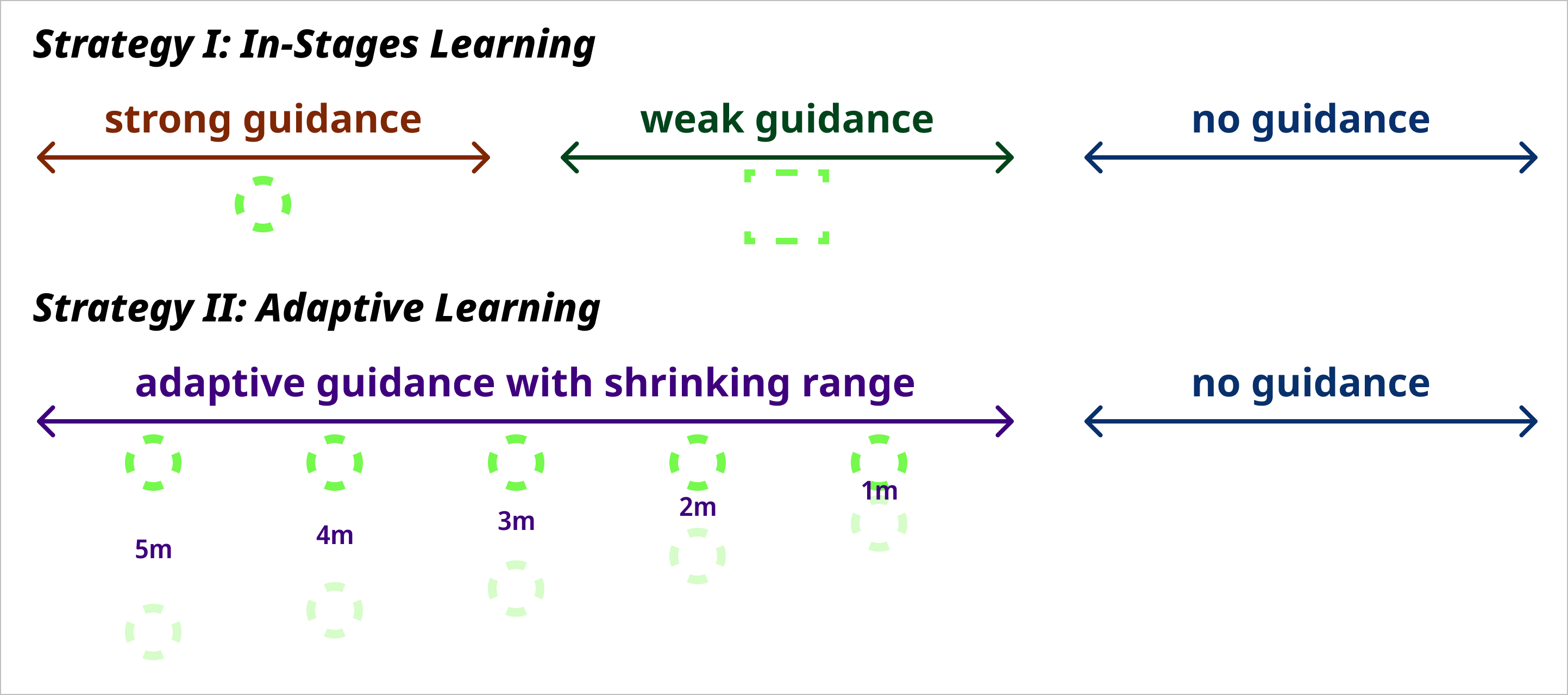}
  \caption{Learning Strategies. In-Stages learning begins with strong guidance at the center of the view, then transfers to weak guidance at the edge, and finally moves to no guidance. Adaptive learning starts with adaptive guidance at the center of the view with variable transparency. As the user progresses, the adaptive range gradually decreases, eventually leading to no guidance.}
  \label{fig:Strategies}
\end{figure}

\section{FocusFlow: Interaction Learning Process}
\label{sec: Learning Strategy}

\revision{Section 5.2 defines several visual cues that could help the users to intuitively shift their gaze depth. However, we argue that depth change is a learnable input method and can become a muscle memory with a proper learning strategy without the need for strong or even any visual cues. To help users develop this new muscle memory, we propose two learning processes based on different series of visual cues: (1) \textbf{in-stages learning} (Figure \ref{fig:Strategies}-I), which walks the user through a sequence of strong to weak visual cues, gradually weakening the cue to remove the user reliance on visual cues for gaze depth transitions. (2) \textbf{adaptive learning} (Figure \ref{fig:Strategies}-II), which leverages the adaptive visual cue with different transparency indexes for the circular visual cue based on the user's visual depth. We describe the technical details of these two learning strategies in the next sections. }

\vspace{.3em}\noindent\revision{\textbf{In-stages Learning}
During in-stages learning process, the UI walks the user through a sequence of gaze-depth interactions with strong, then weak, and finally no visual cues. The learning process mainly relies on repetition and gradually weakening the cues in a plane dimension. In addition a fixed learning sequence is expected to be used for different users. This learning process is simple and efficient but lacks feedback to users on their depth control and perception quality. }

\vspace{.3em}\noindent\revision{\textbf{Adaptive Learning}
The second learning strategy uses the adaptive guidance. At first, a depth range will be defined for activating the visual cue with different transparency indexes assigned to the visual cue based on the user's visual depth. It should be noted that the visual cue's depth range is much larger than the activation zone for virtual window activation. While the adaptive transparency of the visual cue serves as a feedback to the user, we further expand this feedback-based learning process by gradually reducing the depth range of the visual cue until we reach the depth at which the Virtual Window is located. In addition to feedback offering, this learning process can be flexibly modified for different users by changing the rate of visual cue's depth range.}

%% file: tex/6-userstudy.tex
\revision{In this section, we focus on two questions: First, can a gaze-depth input method be intuitively learned to the extent of developing a muscle memory? Second, how efficient is our method and what are the advantages compared with the non-depth hands-free methods? To prove the learnability and efficiency of FocusFlow, we conducted a user study with 24 participants, analyzed the effects of two learning strategies, and compared the performance with the baseline dwell-based selection method as detailed below.}

\subsection{Participants and Apparatus}
\revision{We recruited two groups of participants from the campus with 12 people for each group (Group 1 and Group 2). All of these 24 individuals only have occasional or no VR experience. Out of these 24 individuals, 6 of them reported slight shortsightedness (3 in Group 1 and 3 in Group 2), while 18 of them had a standard uncorrected vision. All the participants did not wear glasses during the test.} In addition, we invited 2 expert users (E1, E2) who had long-term participation in this project as a reference result. The expert users finished all the studies for Group 1 and Group 2. We used the same apparatus as in study \ref{subsec:detection_settings}.

\begin{figure}
 \centering
 \begin{subfigure}[b]{0.22\textwidth}
     \centering
     \includegraphics[width=\textwidth]{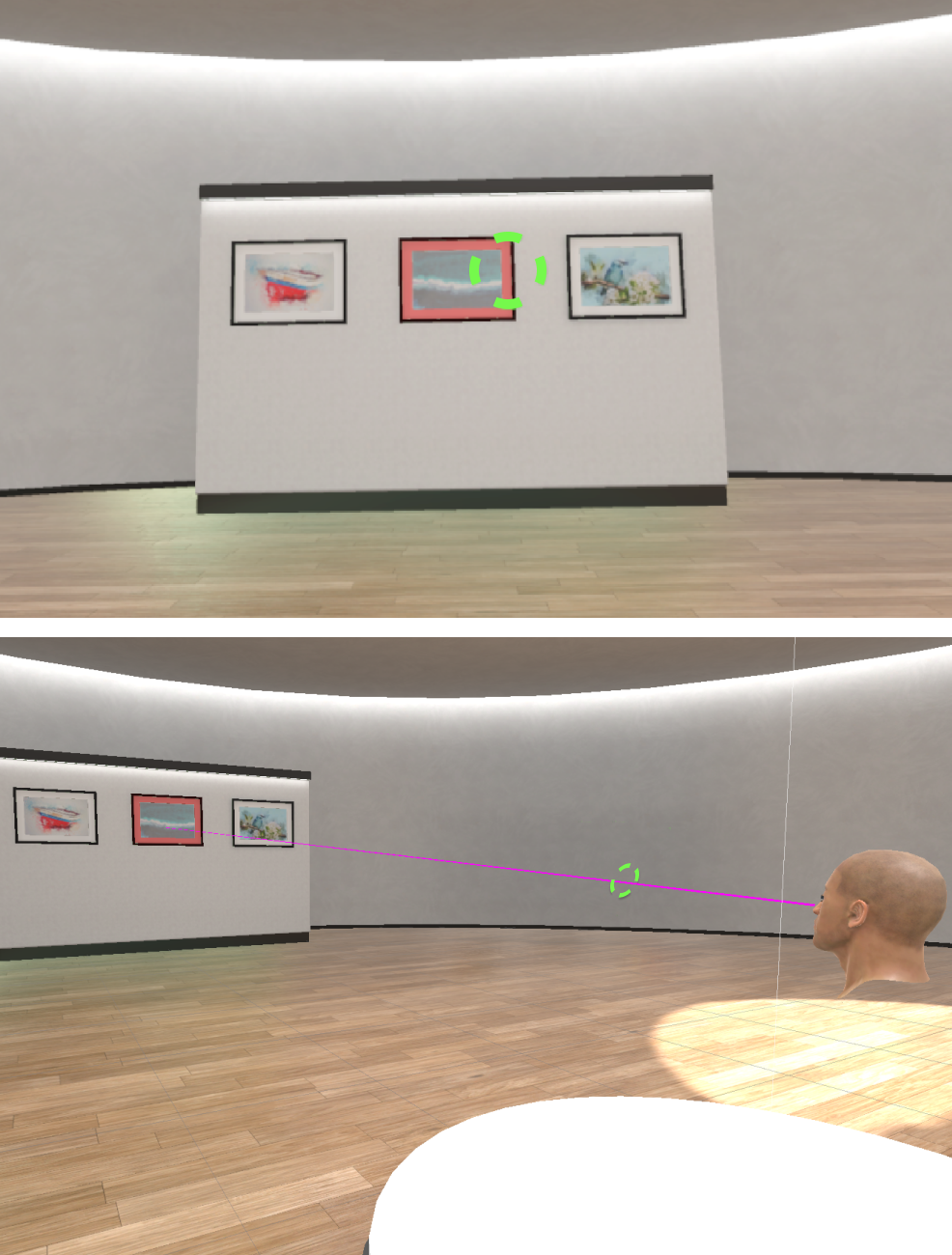}
     \caption{Strong Guidance.}
     \label{fig:user-a}
 \end{subfigure}
 \hfill
 \begin{subfigure}[b]{0.22\textwidth}
     \centering
     \includegraphics[width=\textwidth]{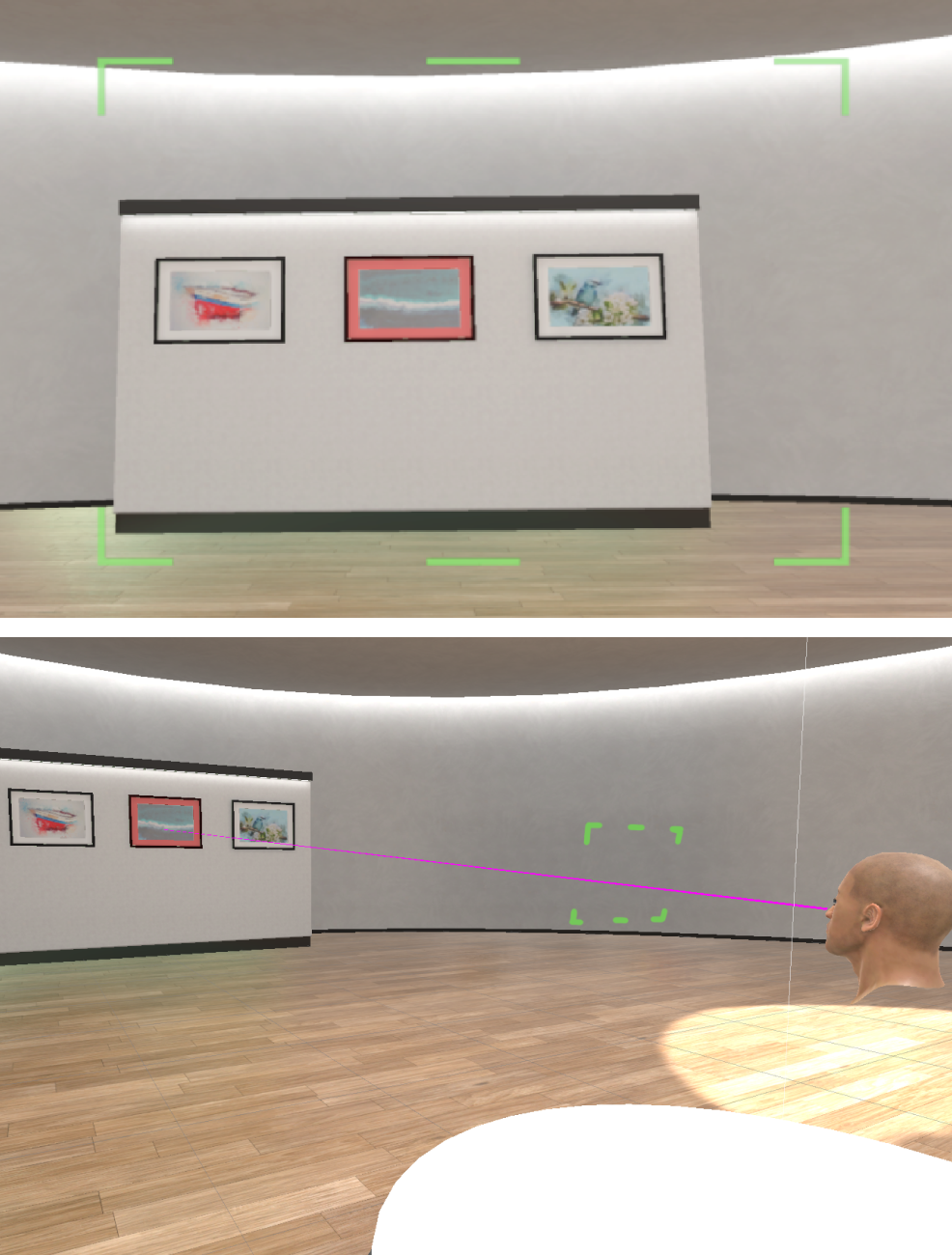}
     \caption{Weak Guidance.}
     \label{fig:user-b}
 \end{subfigure}
 \hfill
 \begin{subfigure}[b]{0.22\textwidth}
     \centering
     \includegraphics[width=\textwidth]{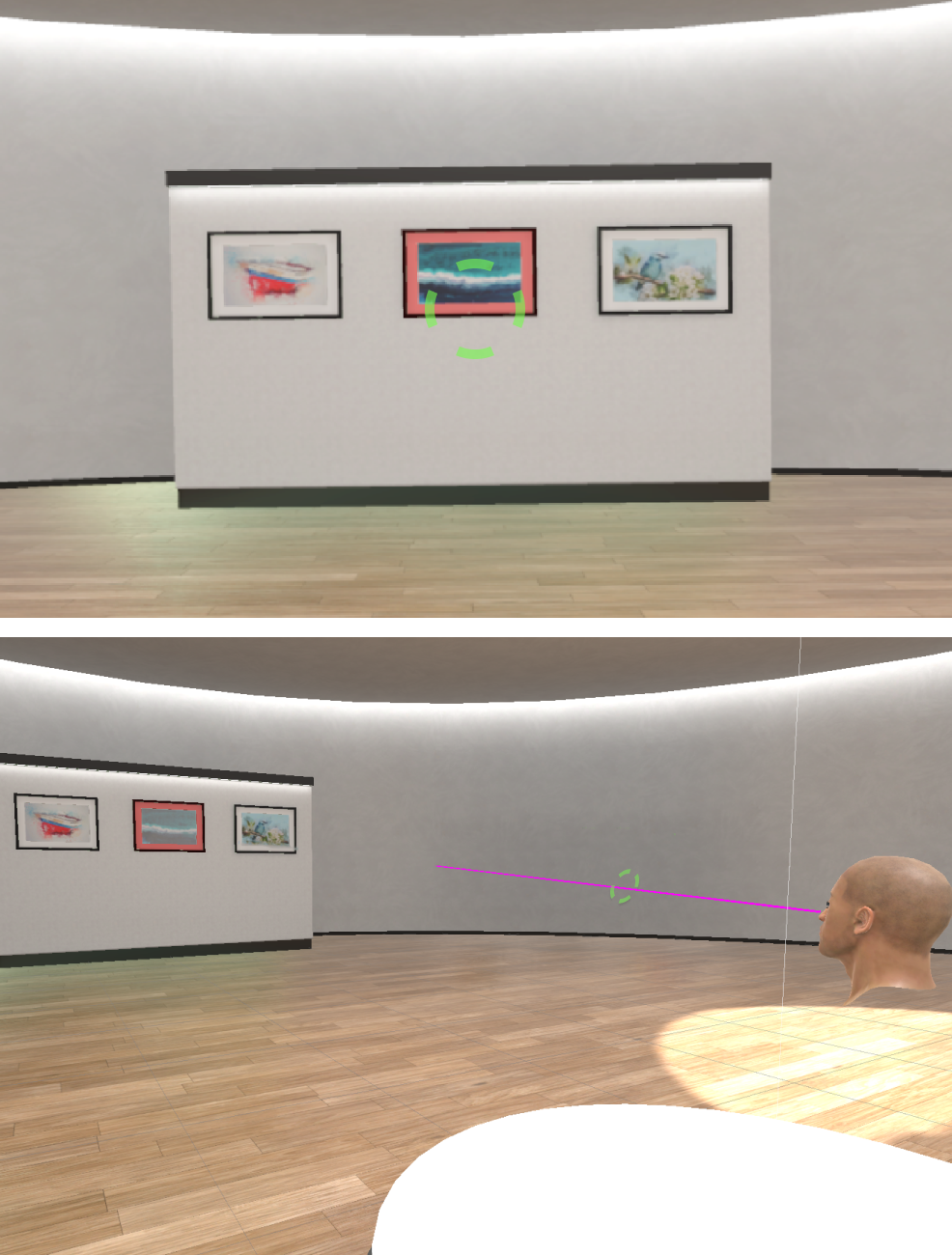}
     \caption{Adaptive Guidance.}
     \label{fig:user-c}
 \end{subfigure}
  \hfill
 \begin{subfigure}[b]{0.22\textwidth}
     \centering
     \includegraphics[width=\textwidth]{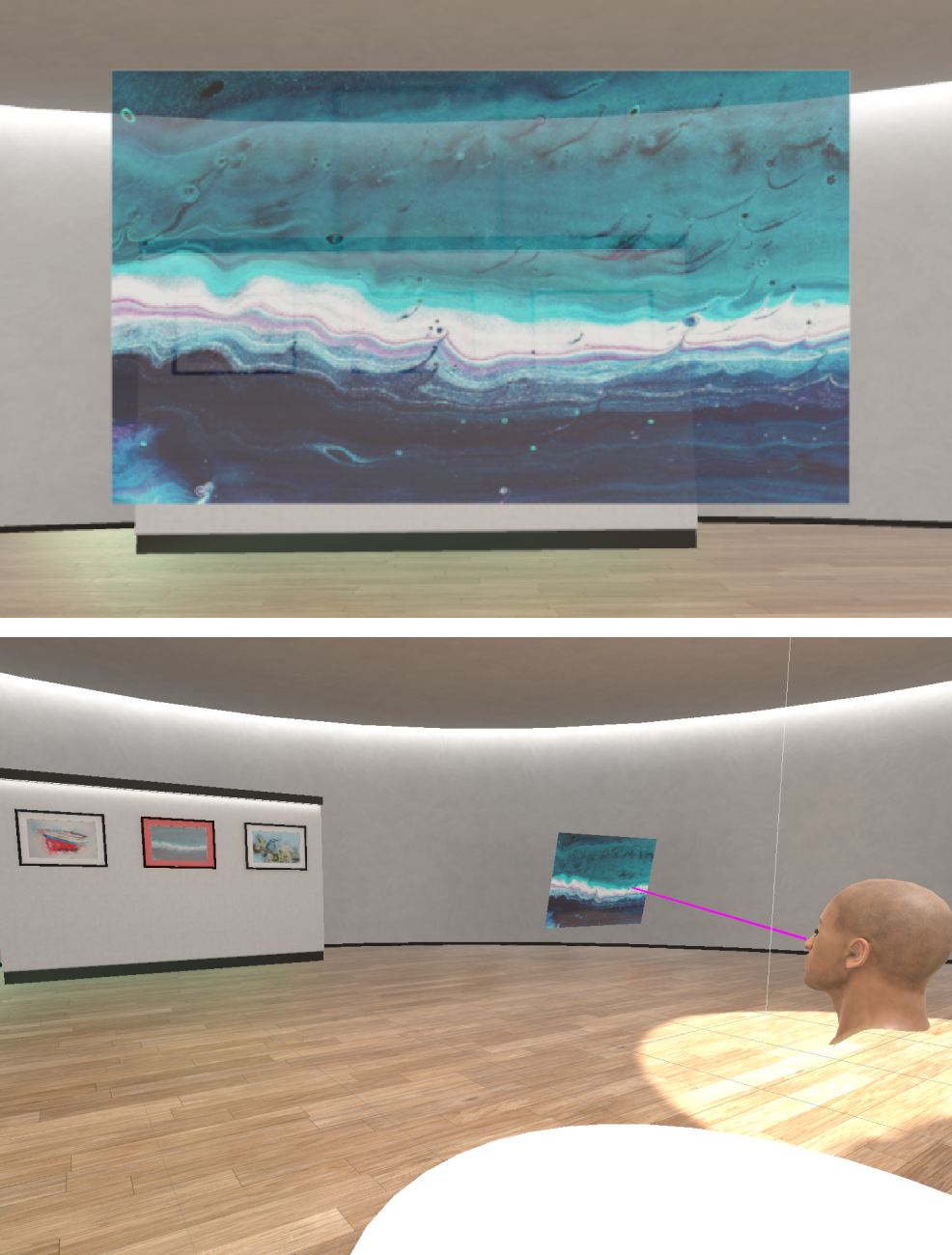}
     \caption{Activation.}
     \label{fig:user-d}
 \end{subfigure}
\caption{User study in a gallery scenario.}
\label{fig:user}
\end{figure}

\begin{figure*}[htb]
\centering
\includegraphics[width=1.0\linewidth]{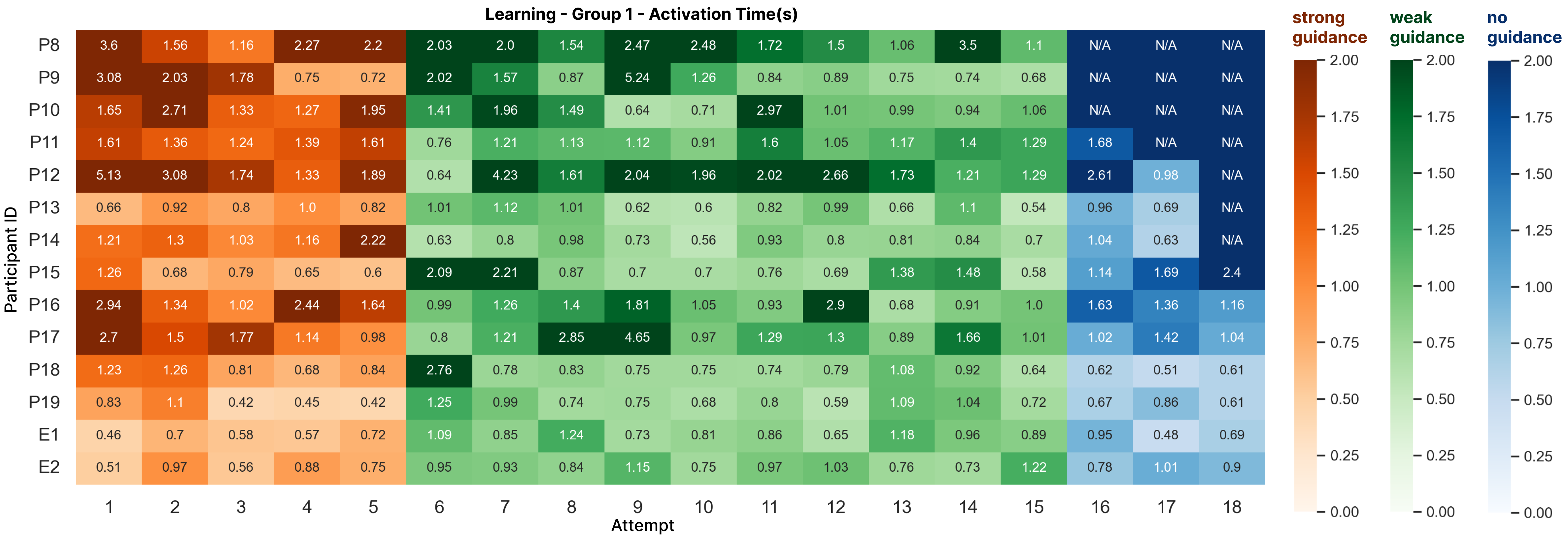}
\caption{Learning attempts of Group 1. Group 1 adopts the in-stages learning strategy with 5 strong-guidance attempts, 10 weak-guidance attempts and 3 no-guidance attempts.}
\label{fig:us_group1_learn}
\end{figure*}

\begin{figure*}[htb]
\centering
\includegraphics[width=1.0\linewidth]{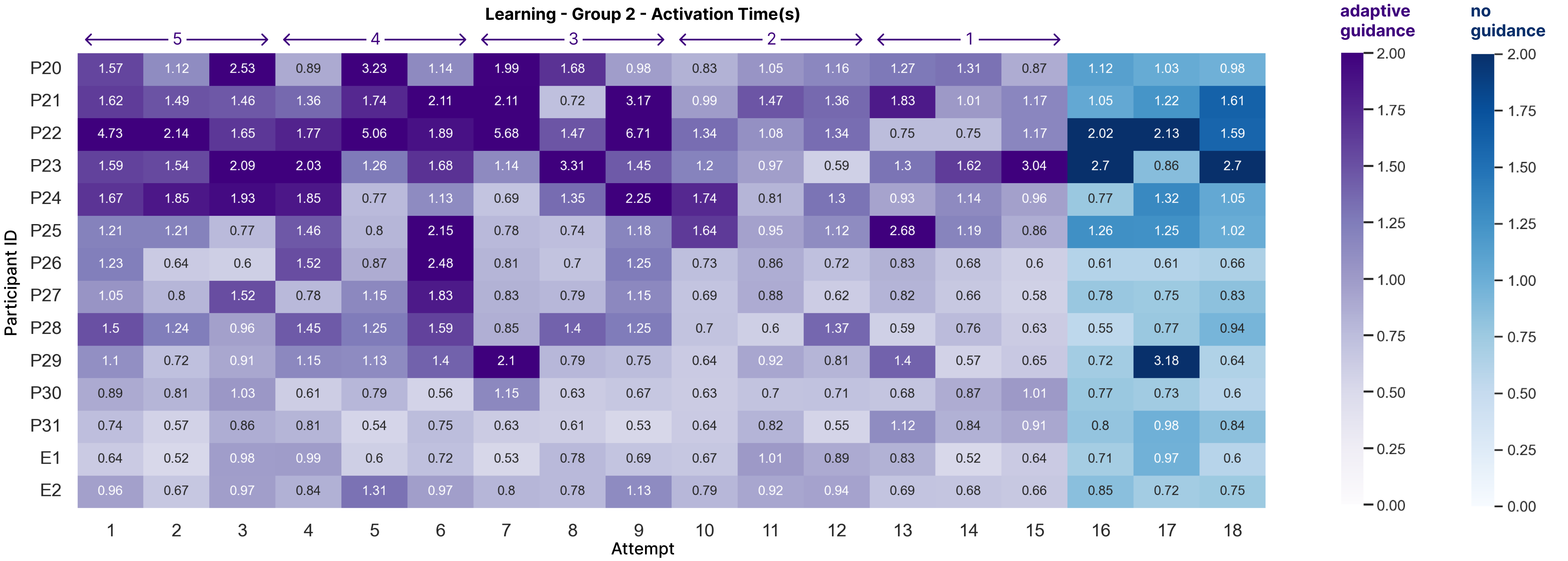}
\caption{Learning attempts of Group 2. Group 2 adopts the adaptive learning strategy with 15 adaptive-guidance attempts and 3 no-guidance attempts. Within the 15 adaptive-guidance attempts, the adaptive range will shrink for every 3 attempts, starting with 5 meters and ending with 1 meter. The target object for selection is 8 meters away from the user.}
\label{fig:us_group2_learn}
\end{figure*}

\subsection{Procedure}
Participants were seated during the experiment, and they could take a rest at any time during the entire process. At first, they put on the headset and performed an eye tracker calibration program. Then they were asked to \textbf{learn} to do gaze selection tasks in a VR gallery shown in Figure \ref{fig:user}, which was to select the target painting to the Virtual Window using the gaze-depth interaction method. This learning process is analyzed in Section \ref{sec: Learning Strategy Analysis}.


\revision{Then they put on the headset again to \textbf{test} their learning effectiveness and proficiency in operation. Participants were first required to finish 10 selection operations within 5 seconds in the same scene, with the help of visual cues. Then they were required to finish another 10 selection operations within 5 seconds without any visual cues. Once they finished the testing stage, the participants will remove the headset and answer the NASA TLX Workload questionnaires to measure their cognitive load.}

Lastly, all the participants in both groups were asked to test the baseline method. They will complete two rounds of 10 selection tasks using the dwell-based gaze interaction method, with time thresholds of 0.5 second and 1 second. Finally, all the participants removed the headset and answered a 7-point Likert scale questionnaire about their user experience. The whole experiment took 20-30 minutes to complete.

\subsection{Learning Strategy Analysis}
\label{sec: Learning Strategy Analysis}

\revision{In this section, we discuss the detailed performance of the two proposed two learning procedures and their learning capabilities by comparing the activation time, users' gaze depth behavior, and the workload during the learning process. With better learning performance and user experience, the adaptive learning strategy is shown as a more effective way for learning gaze-depth interaction than the in-stages learning strategy.}

\subsubsection{\revision{Learning Procedure}}

\revision{Participants in Group 1 are asked to go through the in-stages learning process. As Figure \ref{fig:us_group1_learn} shows, they first do 5 selections with strong guidance, followed by 10 selection attempts with weak guidance, and finally 3 attempts without any visual cues. On the other hand, 
participants in Group 2 are asked to go through the adaptive learning process. As shown in Figure \ref{fig:us_group2_learn}, they perform 15 selection attempts using the adaptive visual cue. The depth range of adaptive guidance start from 5 meters, shrinks 1 meter every 3 attempts, and ends at 1 meter visual depth range which overlaps with the activation zone of Virtual Window. After this adaptive process, the users are asked to perform selections without any visual cue.}

\begin{figure}[htb]
 \centering
 \begin{subfigure}[b]{0.45\textwidth}
     \centering
     \includegraphics[width=\textwidth]{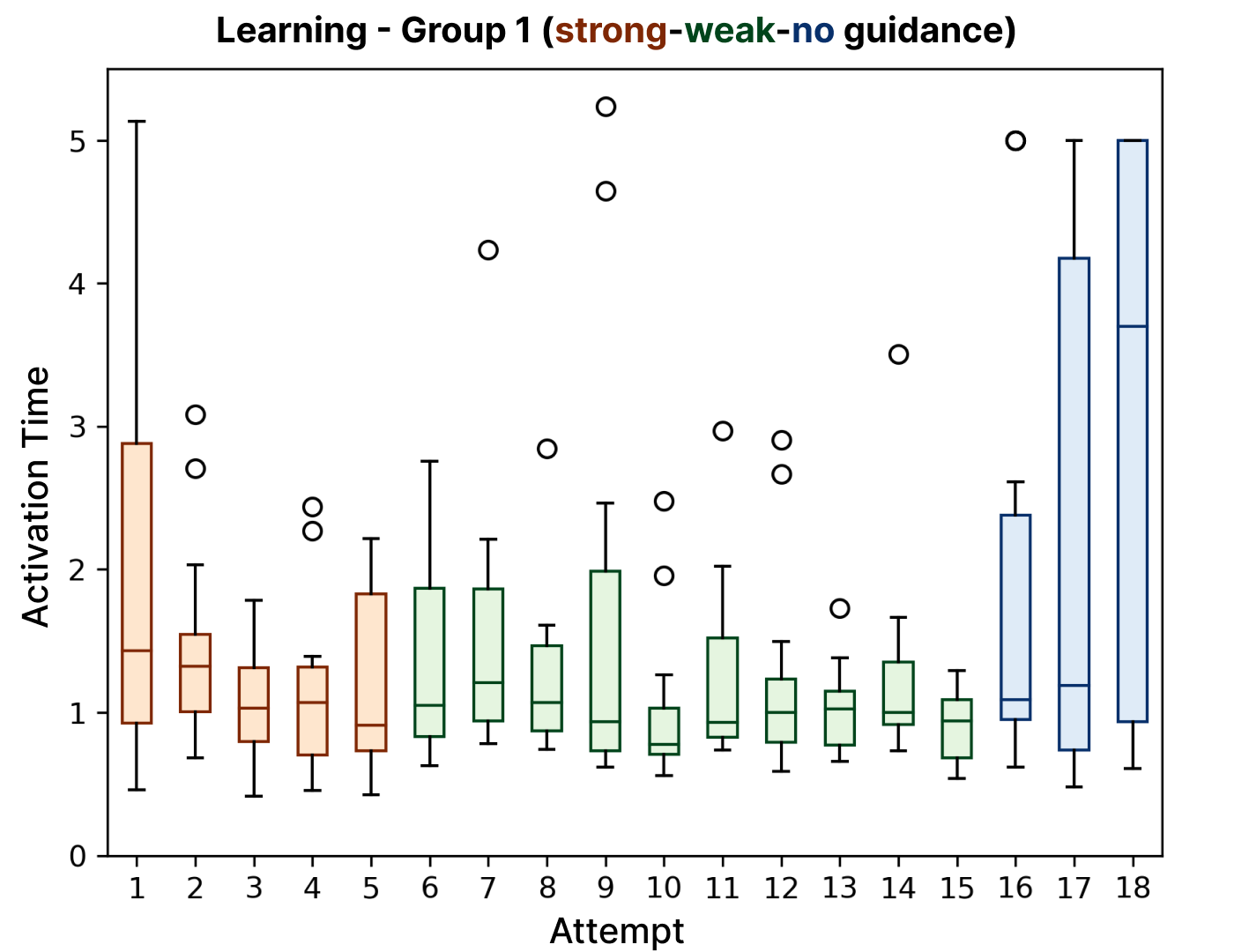}
     \caption{Group 1.}
     \label{fig:box-1}
 \end{subfigure}
 \hspace{0.05\textwidth}
 \begin{subfigure}[b]{0.45\textwidth}
     \centering
     \includegraphics[width=\textwidth]{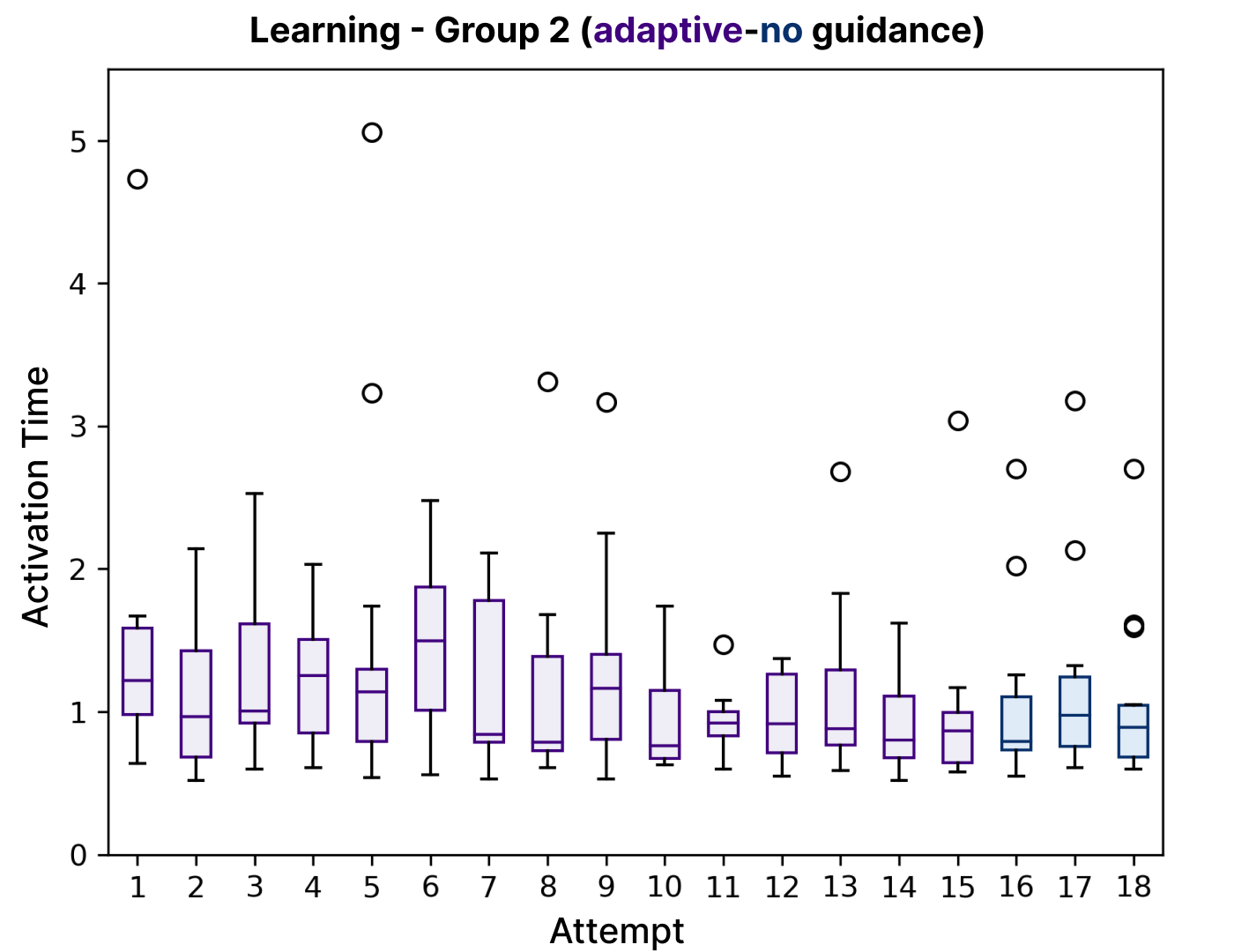}
     \caption{Group 2.}
     \label{fig:box-2}
 \end{subfigure}
\caption{Activation time statistics of all the learning attempts for two groups. the central line in the box indicates the median activation time, while the edges of the box show the interquartile range (IQR), which represents the middle 50\% of the data. Group 1's activation times show a wider range with more variability and outliers, suggesting less consistent performance across attempts. Group 2 displays a more consistent activation time across attempts with fewer outliers, indicating a more uniform performance.}
\label{fig:activation-box-plot}
\end{figure}

\subsubsection{\revision{Activation Time}}
\revision{
We visualize the activation time of all the participant attempts from both groups in the learning stage in Figure \ref{fig:us_group1_learn} and \ref{fig:us_group2_learn}, where the color intensity encodes the activation time. Both figures show a trend of decreasing activation time as learning progresses, demonstrated in lighter colors from left to right for each line/participant. We can also see that 7 participants from Group 1 cannot successfully activate the Virtual Window under no guidance setting, while all the participants from Group 2 can successfully do this. This indicates that learning strategy 2 is more capable of helping users form muscle memory during the learning process.
}

\revision{
We further analyze the effectiveness of learning by comparing the activation time changes in a statistical way. The box plot in Figure \ref{fig:activation-box-plot} shows the distribution of activation times for each round of attempt within two groups. We can see that the activation time of Group 1 (Figure \ref{fig:box-1}) shows a decreasing trend in median activation time specifically  under the strong and weak guidance setting, but this trend does not extend to the no guidance setting. In comparison, Group 2 (Figure \ref{fig:box-2}) shows a consistent decreasing trend under the adaptive guidance and no guidance setting, proving its superiority over learning strategy 1.
}


\subsubsection{Gaze Depth Behavior}

\begin{figure*}[htb]
\centering
\includegraphics[width=1.0\linewidth]{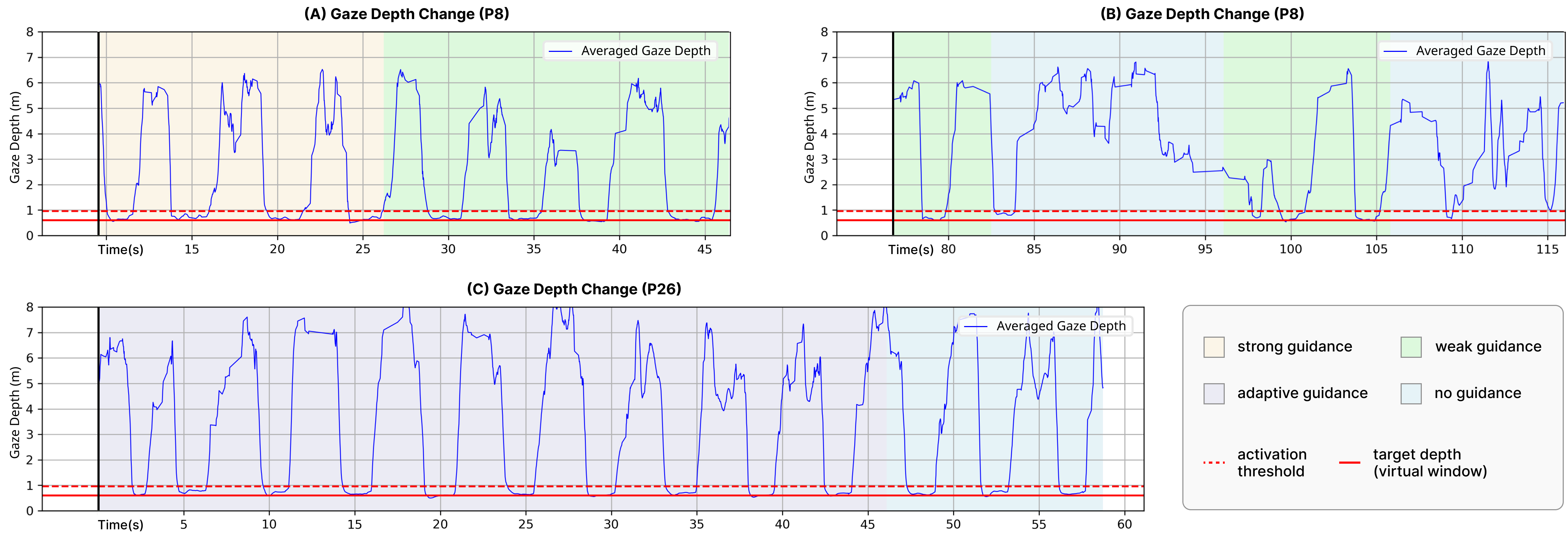}
\caption{Visual Depth Change. The gaze depth change is visualized as the blue line, and the background color indicates different visual cues settings.}
\label{fig:depth-unguide}
\end{figure*}

\revision{
We recorded the gaze depth data during the learning process to investigate the effect of visual cues on the interaction behavior.
Figure \ref{fig:depth-unguide} shows the raw gaze depth values of two representative subjects (P8 and P26).
}

\revision{
P8 adopted the \textit{in-stages learning strategy} and experienced strong guidance, weak guidance, and no guidance activation successively. As Figure \ref{fig:depth-unguide}-A shows, the gaze depth transition of P8 is smooth under the strong guidance and weak guidance setting. However, when entering the no guidance setting from weak guidance (Figure \ref{fig:depth-unguide}-B), P8 made several attempts but can not effectively bring his gaze depth close enough to reach the activation threshold. The reason is it's hard to adjust the gaze depth accurately without any depth referent or muscle memory. The in-stages learning strategy fails to help P8 to develop the muscle memory, so as the system switched to the no guidance setting and there was nothing to be a referent, P8 can not shift his gaze depth to the target distance.
}

\revision{
The \textit{adaptive learning strategy} shows a better effectiveness on the transition from adaptive guidance to no guidance. As Figure \ref{fig:depth-unguide}-C shows, P26 can still stably shift the gaze depth between close distance and far distance, proving the feasibility of forming a muscle memory. This is because the shrinking assisted range is continuously reduce users' dependence on visual cues, and the changing opacity provides a clear feedback to users to help them understand their input status.
}


\begin{figure}
 \centering
 \begin{subfigure}[b]{0.45\textwidth}
     \centering
     \includegraphics[width=\textwidth]{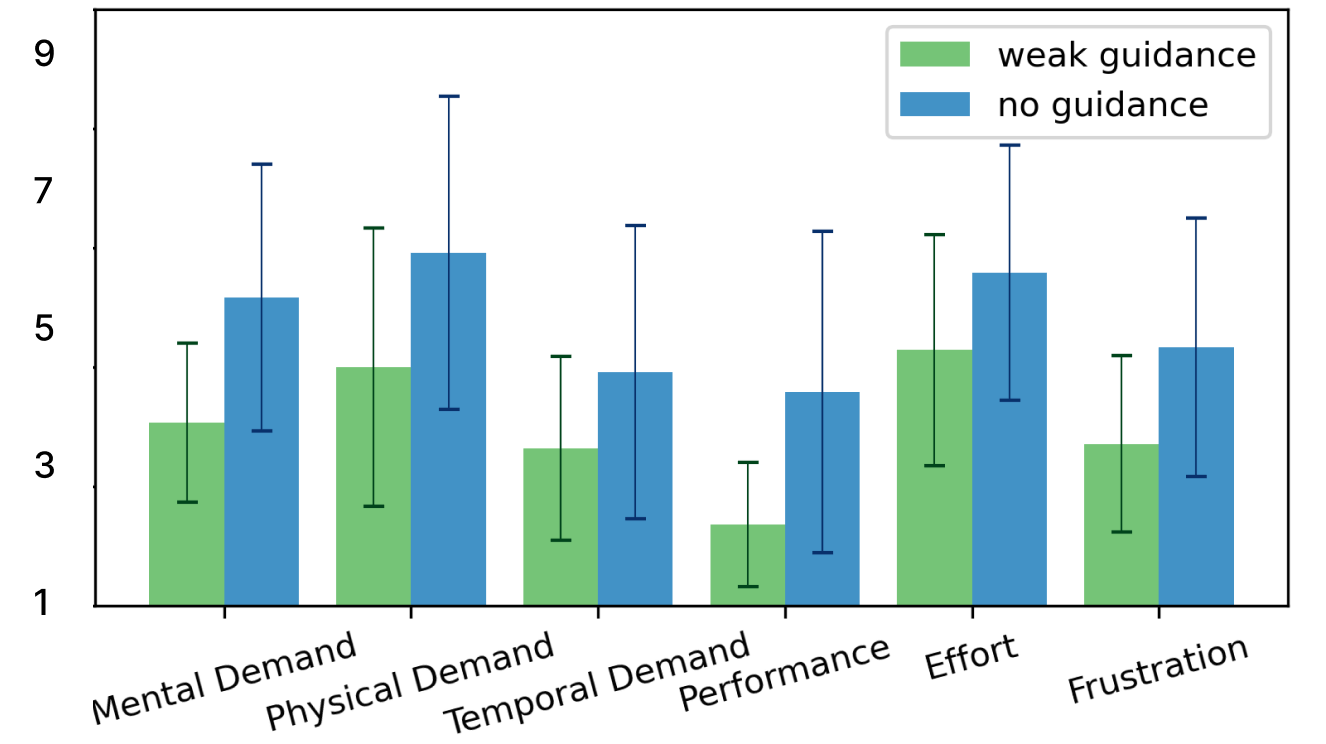}
     \caption{Group 1.}
     \label{fig:NASA-TLX G1}
 \end{subfigure}
 \hfill
 \begin{subfigure}[b]{0.45\textwidth}
     \centering
     \includegraphics[width=\textwidth]{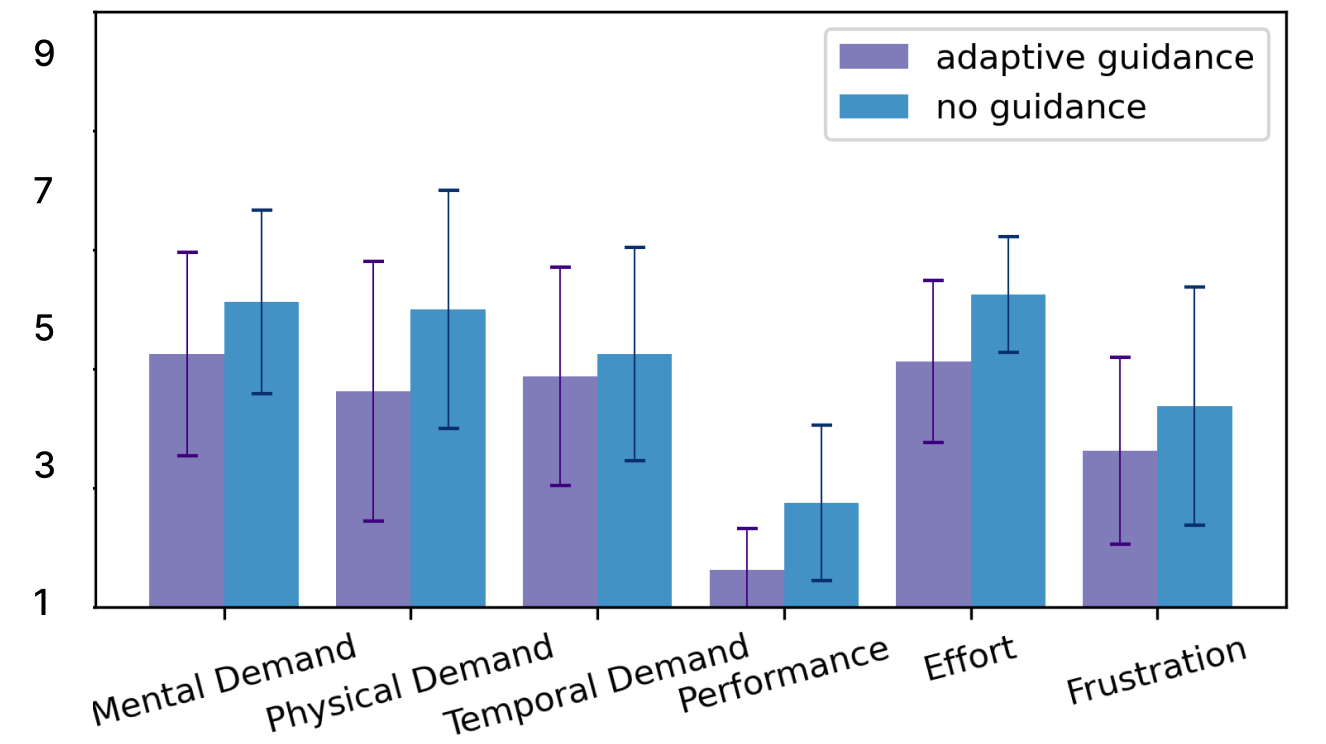}
     \caption{Group 2.}
     \label{fig:NASA-TLX G2}
 \end{subfigure}
\caption{NASA-TLX Results. We measure the subjective workload during two learning process (in-stages learning and adaptive learning) using the NASA-TLX Index. For each entry, lower score means lower workload and better experience.}
\label{fig:NASA-TLX}
\end{figure}


\subsubsection{Workload}

\revision{
We employed the NASA Task Load Index (NASA-TLX) to assess the cognitive and physical workload of two learning processes. The result in Figure \ref{fig:NASA-TLX} shows that the no guidance method imposes higher cognitive and physical demands on the users than the guided methods, indicating the muscle memory development requires some effort. In addition, the cognitive load scores of adaptive guidance is slightly higher than the weak guidance method, because its changing capacity brings more information to users to process. However, it also shows a lower performance load score (lower is better), which means eventually the adaptive guidance can offer better activation performance.
}


\begin{figure}[htb]
 \centering
 \begin{subfigure}[b]{0.48\textwidth}
     \centering
     \includegraphics[width=\textwidth]{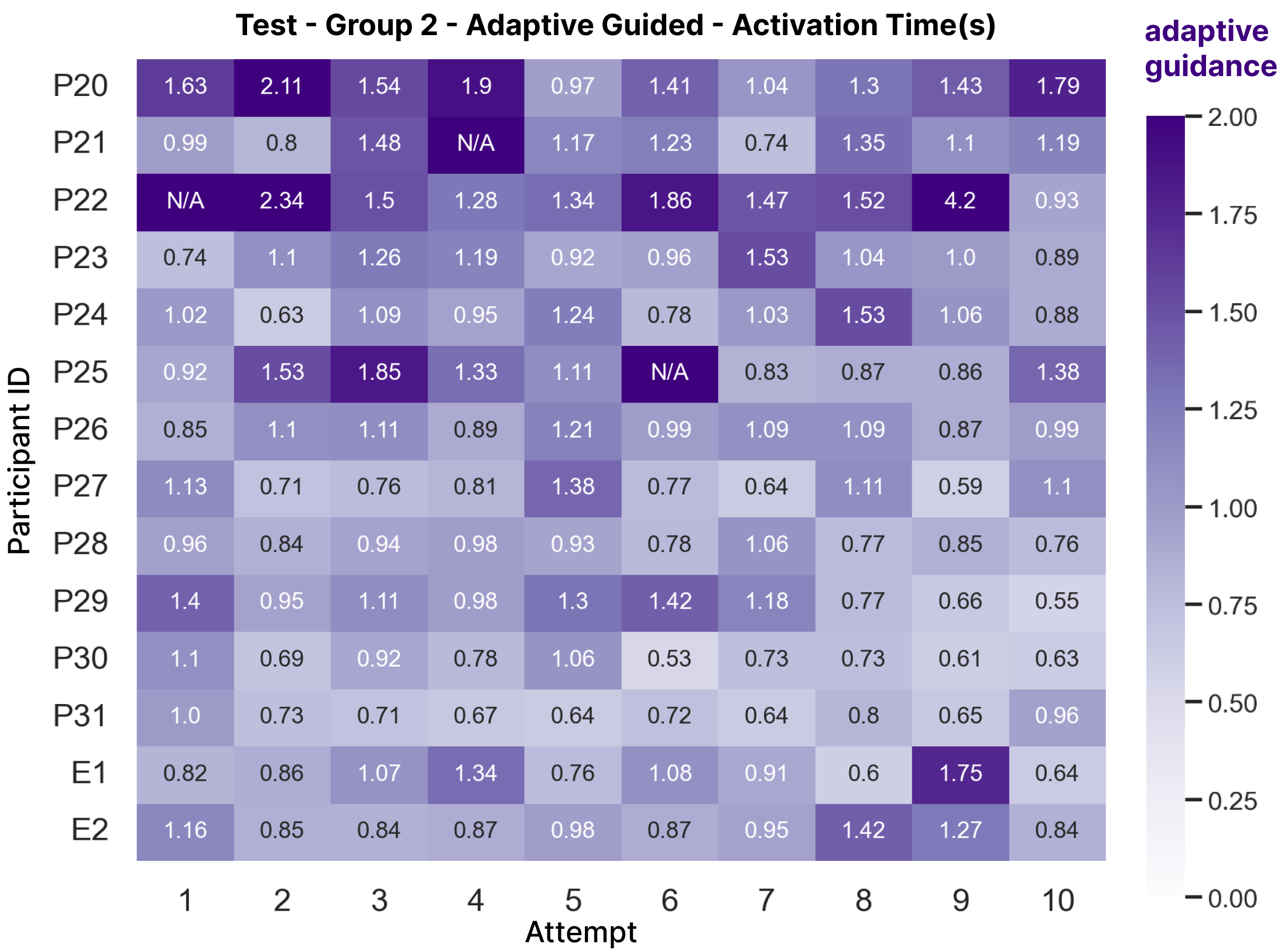}
     \caption{Test with adaptive guidance of Group 2.}
     \label{fig:test-adaptive}
 \end{subfigure}
 \hfill
 \begin{subfigure}[b]{0.48\textwidth}
     \centering
     \includegraphics[width=\textwidth]{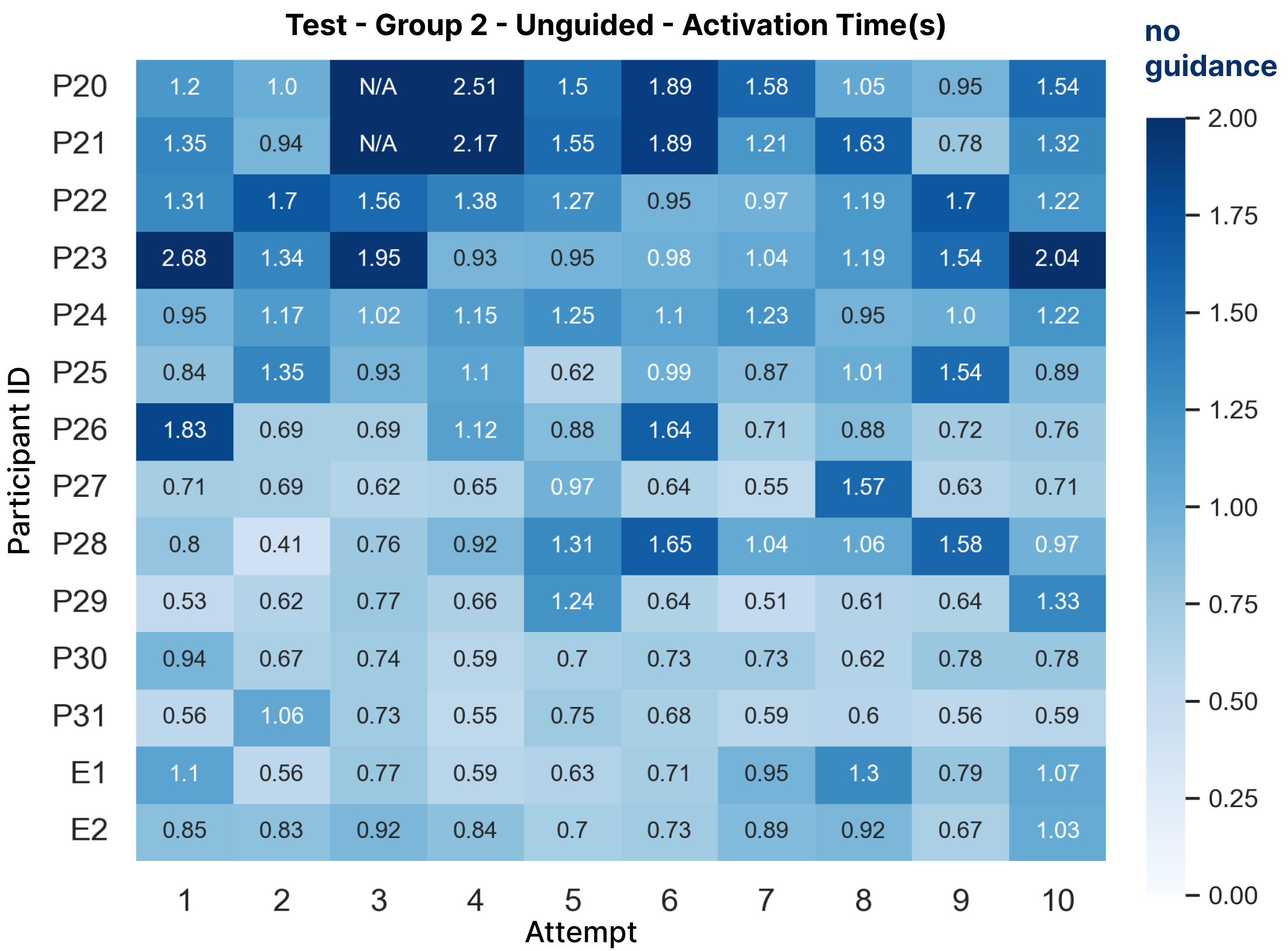}
     \caption{Test with no guidance of Group 2.}
     \label{fig:test-no}
 \end{subfigure}
\caption{Activation time for the tests of Group 2 in two guidance settings. In the test with no guidance, users in Group 2 can still finish most of the selection tasks in a similar activation time to the time used in the test with adaptive guidance.}
\label{fig:test-group2}
\end{figure}

\subsection{Performance Analysis}
\label{sec: Performance Analysis}

\begin{figure*}[htb]
\centering
\includegraphics[width=1.0\linewidth]{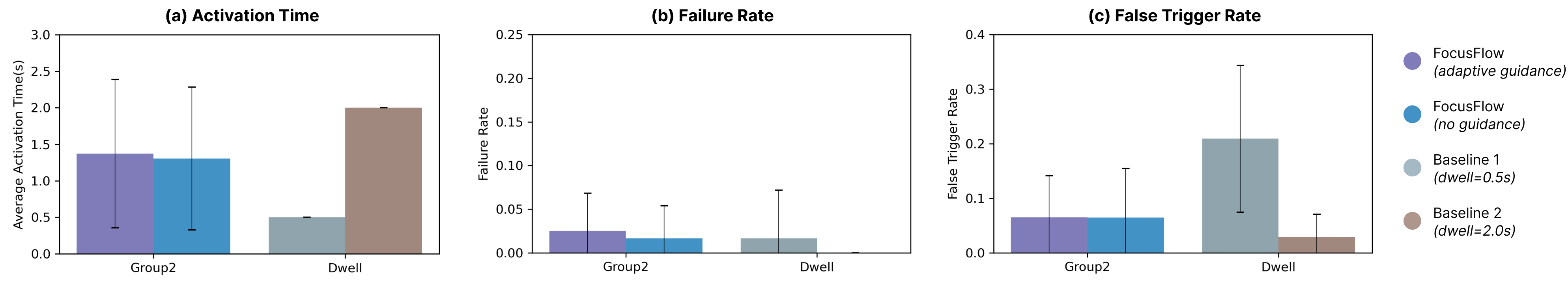}
\caption{Performance comparison between FocusFlow and dwell methods in different settings.}
\label{fig:performance}
\end{figure*}

\revision{The analysis in Section \ref{sec: Learning Strategy Analysis} revealed that the adaptive learning strategy enhances participants' proficiency with the gaze-depth interaction method in FocusFlow. Building on this insight, this section delves into the performance of Group 2 during their test stages, comparing it to the baseline dwell method to assess the efficiency and accuracy of FocusFlow.}

\subsubsection{Activation Time}
\revision{We define 'activation time' as the duration from the moment a user's gaze targets an object to when the virtual window appears. Figure \ref{fig:test-adaptive} and \ref{fig:test-no} illustrate the activation times across all selection attempts by Group 2 during two tests in the testing stage. It's important to note that dark squares marked as 'N/A' represent instances where participants didn't complete the selection task within the 5-second limit. Remarkably, participants in Group 2 successfully completed the selection tasks in the no-guidance test setting. This achievement underscores their high proficiency with the gaze-depth interaction method, particularly when contrasted with the last several attempts shown in Figure \ref{fig:us_group1_learn} for the in-stage learning strategy. Additionally, their activation times in the no-guidance test are comparable to the time in the adaptive-guidance test, suggesting that Group 2 participants have effectively mastered the gaze-depth interaction method without any visual guidance. Moreover, we also compare the activation time between the gaze-depth method (FocusFlow) and the baseline dwell method in Figure \ref{fig:performance}a. We could find that the two settings in FocusFlow achieve a similar activation time of around 1.3s, which is shorter than the 2-second dwell method, while the half-second dwell method achieved the shortest time.}

\subsubsection{Failure Rate}
\revision{We define 'failure rate' as the ratio of successful selection attempts to the total number of attempts in each stage, considering an attempt a failure if it's not completed within the time limit. Figure \ref{fig:performance}b reveals that all four settings exhibit a failure rate below 5\%, which suggests that the reliability of FocusFlow is comparable to that of the traditional dwell method.}

\subsubsection{False Trigger Rate}
\revision{We define the 'false trigger rate' as the proportion of incorrect selections relative to the total number of selections, with 'incorrect selections' being those where the wrong targets are chosen. As illustrated in Figure \ref{fig:performance}c, the half-second dwell method exhibits the highest false trigger rate. This may be attributed to the method's overly sensitive nature, leading to an increased likelihood of unintentional target activations. In contrast, both settings in FocusFlow demonstrate a considerably lower false trigger rate, approximately 5\%. This rate is marginally higher than that observed in the 2-second dwell method.}

\subsubsection{Overall Performance}
\revision{In the preceding paragraphs, we evaluated various metrics of FocusFlow, comparing them against two baseline dwell time settings as depicted in Figure \ref{fig:performance}. Overall, FocusFlow demonstrates commendable performance with an average activation time of 1.3 seconds, a failure rate under 5\%, and a relatively low false trigger rate of approximately 5\%. While the half-second dwell method offers a shorter average activation time and a comparably low failure rate, its high incidence of false triggers could negatively impact user experience. Conversely, the 2-second dwell method, despite its stability and low rates of failure and false triggers, may lead to user frustration due to longer activation times, especially in certain scenarios. Additionally, the dwell method restricts users from continuously observing the target object, which might detract from the overall user experience. Thus, FocusFlow is a reliable and efficient interaction choice with an appropriate learning strategy applied to enhance the user experience in the VR space.}

\begin{figure*}[htb]
\centering
\includegraphics[width=1.0\linewidth]{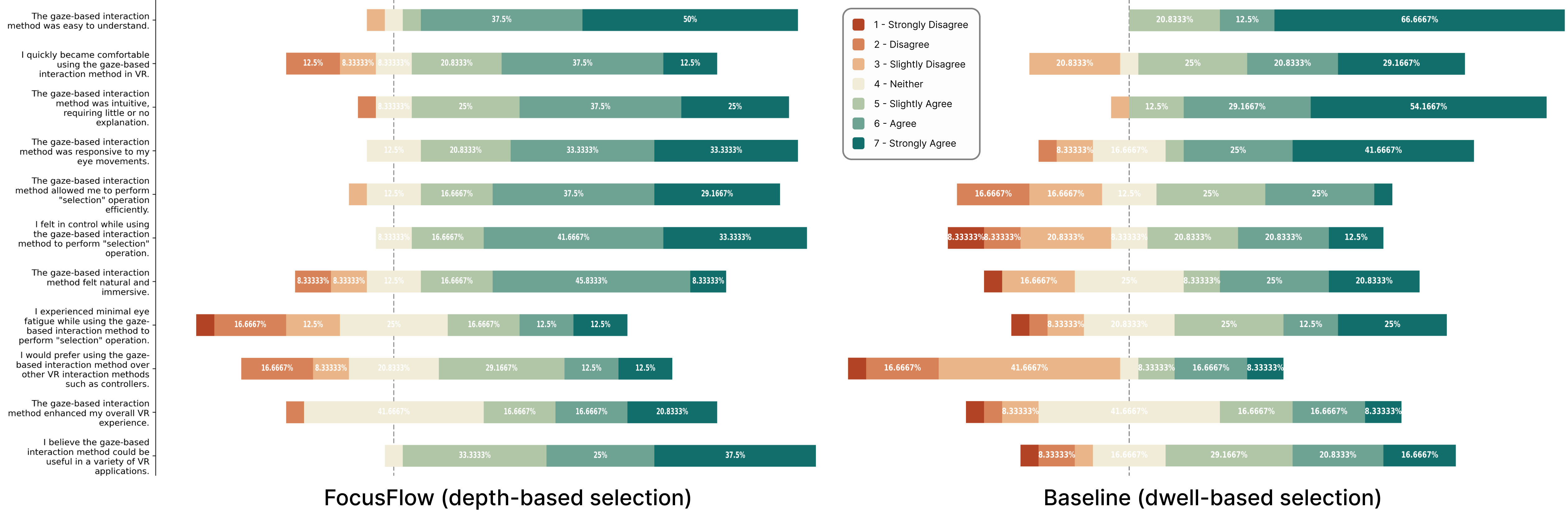}
\caption{User Experience Likert Data.}
\label{fig:likert}
\end{figure*}

\subsubsection{User Experience}

Besides the quantitative metrics we discussed above, we also designed a Likert questionnaire to evaluate the user experience of FocusFlow, focusing on its usability, efficiency, comfort, overall preference, and potential applicability in various VR contexts, and compared the results with the dwell-based baseline method. The result in Figure \ref{fig:likert} shows that FocusFlow receives positive feedback from participants, which proves its usability and learnability.
Compared to the baseline method, FocusFlow is overall better and obviously outperforms it in terms of efficiency (Q.4, Q.5, Q.6) and overall experience (Q.9, Q.10). 
It also shows that the eye fatigue when using FocusFlow is more serious than the baseline dwell-based method, which should be an issue that needs to be emphasized and improved upon.
In addition, users are wary of using hands-free gaze interaction as an alternative to other VR interaction methods. This suggests that gaze-only interaction cannot cope with all VR interaction scenarios, so compatibility and extendibility with other interaction methods should also be an important design consideration.

%% file: tex/7-casestudy.tex
\begin{figure}
 \centering
 \begin{subfigure}[b]{0.23\textwidth}
     \centering
     \includegraphics[width=\textwidth]{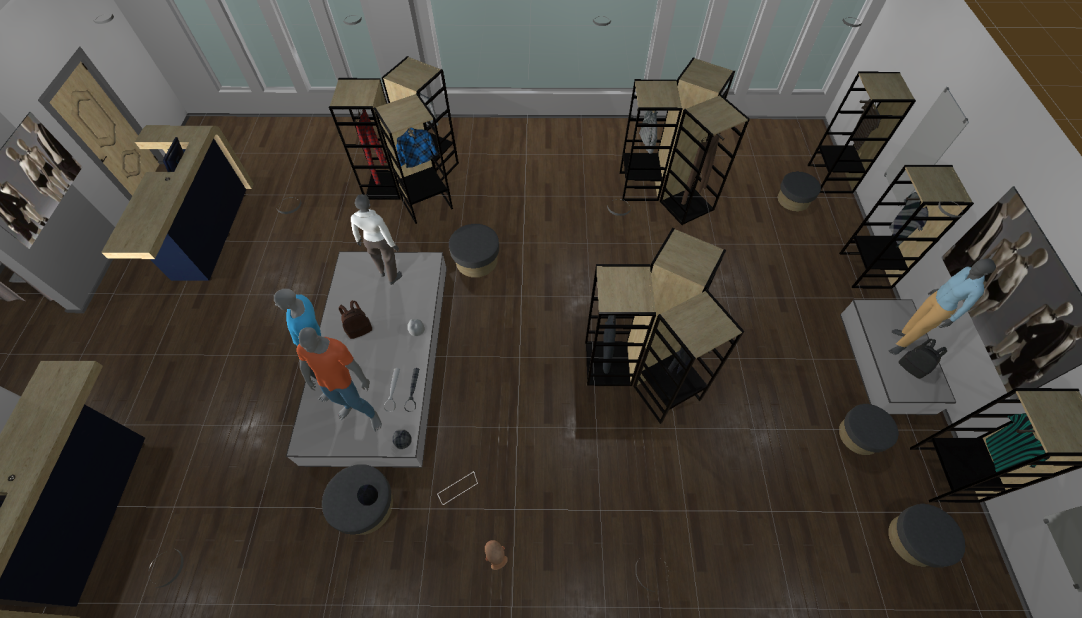}
     \caption{Scene Overview}
     \label{fig:case-a}
 \end{subfigure}
 \hfill
 \begin{subfigure}[b]{0.23\textwidth}
     \centering
     \includegraphics[width=\textwidth]{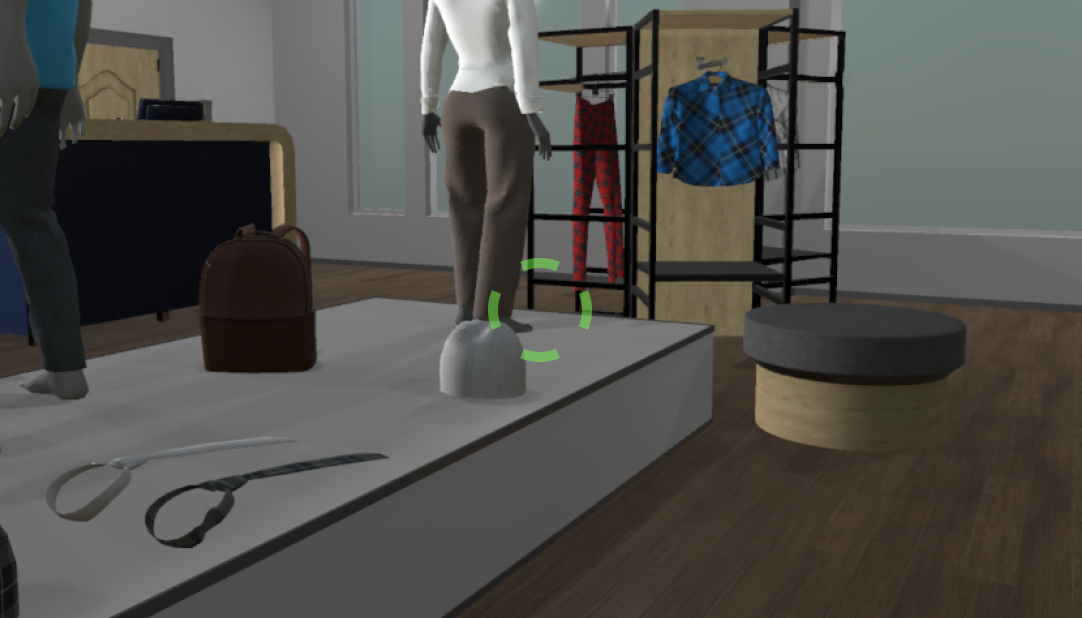}
     \caption{Visual Cues}
     \label{fig:case-b}
 \end{subfigure}
 \hfill
 \begin{subfigure}[b]{0.23\textwidth}
     \centering
     \includegraphics[width=\textwidth]{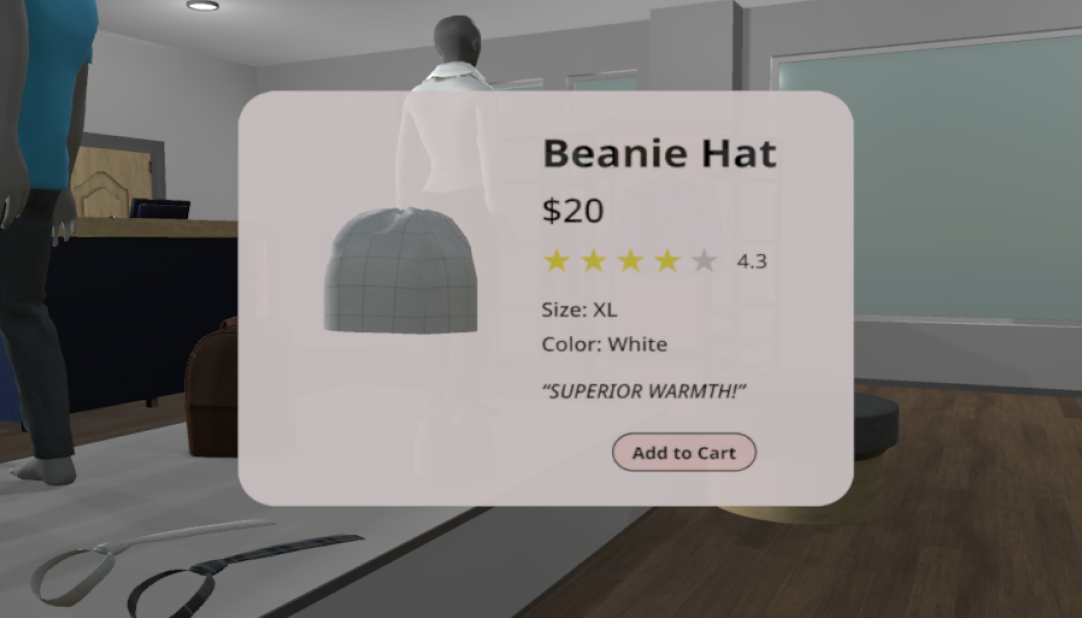}
     \caption{Quick Preview Window}
     \label{fig:case-c}
 \end{subfigure}
\caption{Case study in a shopping scenario.}
\label{fig:application}
\end{figure}

FocusFlow facilitates intuitive and accurate hands-free selections in VR. To demonstrate how users interact with FocusFlow and their perceptions of the process, we present a case study set in a virtual fashion store shopping environment that mimics real-world scenarios. (Figure \ref{fig:case-a})

In this environment, each product can trigger a preview window that showcases its profile. This is consistent with the Virtual Window setting encountered during user training. Items within the store vary in distance from the user, ranging between 2 to 7 meters. Users are encouraged to freely explore these products, mirroring a genuine shopping experience. The dwell method and our depth-based method are employed successively to ascertain the experiential difference conferred by the addition of a depth input dimension.

Utilizing \textbf{FocusFlow}, users reported a seamless and efficient experience. Initially, they would pan their heads around to acquaint themselves with the virtual space. Gradually, their attention would focus on specific items of interest. At this juncture, visual cues become evident (Figure \ref{fig:case-b}).
\userquote{The appearance of adaptive visual cues hinted at the target depth and gave me some feedback on my gaze depth, allowing me to confidently engage in depth adjustments to activate the virtual window.}
\userquote{I didn't experience significant visual disruption since the depth of the adaptive visual cues didn't match the depth of the object I was observing. Thus, I could easily overlook its presence when focused on an item.} (P19)
To access detailed information about an item, users adjusted their gaze to the depth of the virtual window (Figure \ref{fig:case-c}). This action promptly displayed the product details. Subsequently, by shifting their focus, users could effortlessly continue to survey the environment.
\userquote{The activation and deactivation mechanism was very fluid. It felt as though a virtual window was consistently present; and whenever I looked at it, information would display.} (P8)

However, during free exploration using the \textbf{dwell method}, users faced considerable challenges. When their gaze lingered on a product, the sudden appearance of the information panel – triggered by surpassing the observation time threshold – obstructed their view.
\userquote{The most significant flaw of this method is that I can't freely gaze at an object indefinitely. I constantly have to be cautious to not stare too long, lest the window pops up, hindering my observation. I believe the freedom to observe the world undisturbed is crucial for user experience.} (P21)
The lag between intention to inspect and the system's response was also disruptive.
\userquote{During that waiting period, even if it's just a second, I felt anxious. It felt like wasted time since the system didn't promptly capture my intent.} (P11)
After inspecting a product, to exit the information panel, users had to move their gaze to its periphery, an action they deemed superfluous.
\userquote{The dwell method required me to shift my gaze to the edge to exit. There was a continual need to oscillate between the center and periphery. In contrast, with the depth method, I merely had to naturally look past the translucent panel towards the distant environment to exit. Essentially, with the depth method, I wasn't burdened with unnecessary actions since my goal was to resume observation, which in itself served as the exit mechanism.} (P9)

In summary, users welcomed the depth input dimension.
\userquote{The depth method gave me a stronger sense of control. I could distinctly decide whether to activate a target.} (P6)
\userquote{I preferred the depth method. Though time-based techniques are simpler, once I got the hang of the depth method, I noticed an improved experience.} (P8)
However, they also highlighted that the depth method was more straining on the eyes after frequent use.
\userquote{Compared to the dwell method, a downside of the depth approach is the fatigue it induces in my eyes due to the constant need for depth adjustments.} (P13)
This feedback implies that the depth method might not be ideal for extended use in short durations, as continuous adjustments could tire the eyes.

%% file: tex/8-discussion.tex
FocusFlow features a hands-free gaze-depth-based input method that is efficient and intuitive for selection tasks after a minimal learning process. We proposed a layer-based user interface design according to the visual depth estimation characteristics and learning strategies to help users learn this gaze-depth input method. 
 Although our evaluation results demonstrate FocusFlow is superior to baseline in terms of operational efficiency and the effectiveness of the adaptive learning strategy in helping users master this gaze-depth input method, there are still some aspects that could be improved or extended in future works:

\vspace{.3em}\noindent\textbf{Statistical Analysis:}
\camready{ The current paper does not include any major statistical significance analysis due to small participant population. In our future work, we plan to perform a long-term experiment with a larger user population to fully demonstrate the effectiveness and intuitiveness of the proposed approach. 
}



\vspace{.3em}\noindent\textbf{Multi-modal Interaction Inputs:}
\camready{
Visual depth is a unique input dimension with a marginal likelihood of clashing with other eye behaviors. However, it only offers limited interaction inputs such as selection or zoom-in. To improve the number of inputs, FocusFlow can be seamlessly integrated with other eye input methods, such as eyelid movements or blinks. Moreover, the visual depth modality can be harmoniously coupled with other VR interaction modes like hand and head inputs. For instance, a user might activate a virtual window through gaze-depth interaction and subsequently interact with components within the window using direct or indirect hand gestures \cite{2017pfeufferphd}. Such compound interactions can be explored in detail in future research.
}

\vspace{.3em}\noindent\textbf{Gaze Depth as a Continuous Input:}
\label{fine grained input}
\camready{The evaluations in Section \ref{detection_analysis} show that high-accuracy depth estimations are limited to close distances. While this limits our accurate depth detection over long distances, we can still utilize depth data at close range for more fine-grained inputs. For example, we can expand our gaze-depth interaction from binary selection to continuous inputs by continuously tracking exact visual depth changes and mapping the depth to some operational values. This will enable various manipulations such as scrolling on a webpage \cite{sharmin2013reading} or rotating a 3D object \cite{liu2020eye} by looking at micro gaze-depth changes.}

\vspace{.3em}\noindent\textbf{Virtual Window Depth Adaptability:}
\revision{To elevate user experience and reduce erroneous virtual window activation, the position of the interaction layer could be adapted to a target's location. For example, when the target is a few meters away from the user, the interaction layer could be set to one meter in front of the user. However, if the target distance is less than one meter, the position of the interaction layer will be automatically set to half a meter from the user. However, such adaptability necessitates visual cues to inform users of the adjusted window position, ensuring seamless interactions.}

\begin{figure}
 \centering
 \begin{subfigure}[b]{0.45\textwidth}
     \centering
     \includegraphics[width=\textwidth]{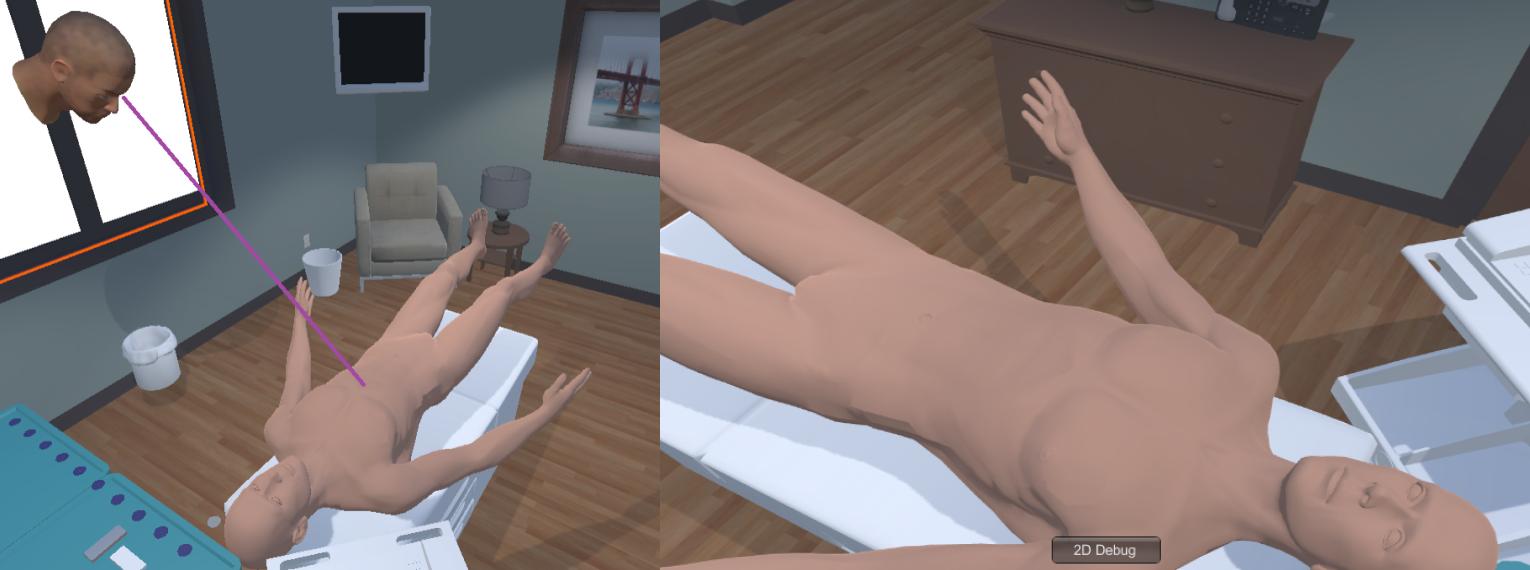}
     \caption{Looking at the object}
     \label{fig:Surgical-a}
 \end{subfigure}
 \hfill
 \begin{subfigure}[b]{0.45\textwidth}
     \centering
     \includegraphics[width=\textwidth]{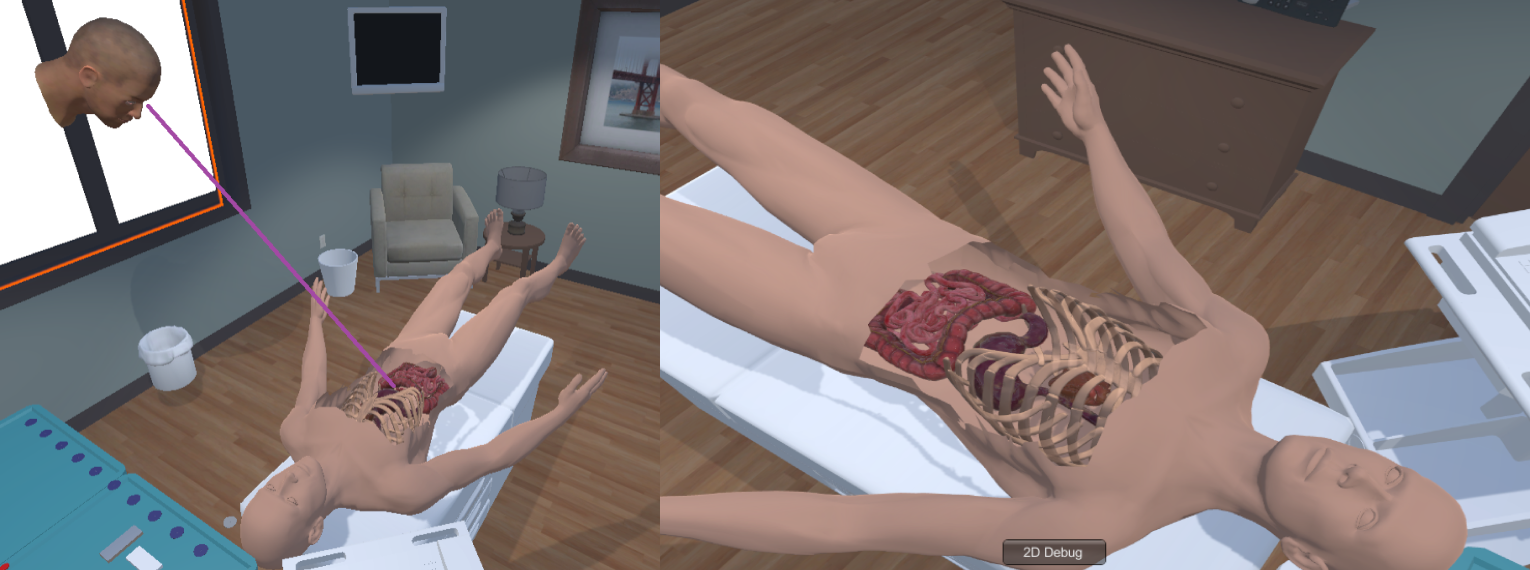}
     \caption{Looking through the object}
     \label{fig:Surgical-b}
 \end{subfigure}
\caption{Surgical training application.}
\label{fig:Surgical application}
\end{figure}

\vspace{.3em}\noindent\textbf{Looking Through Interaction:}
\camready{
In this paper, we assumed that the virtual window is always placed closer to the user than the interactive object. However, FocusFlow can be seamlessly extended to applications where information is designed further from the interactive object. For example, in the surgical training application shown in Figure \ref{fig:Surgical application}, the user can switch between observing different layers inside the patient's body by actively changing their focal depth further from the upper surface of the body.
}

%% file: tex/9-conclusion.tex
We present \emph{FocusFlow}, a novel gaze-depth interaction technique that leverages the potential of visual depth as an intuitive input dimension. Our approach introduces a layer-based user interface that resonates with the concept of a ``Virtual Window'', offering an efficient and hands-free selection method. By doing so, we address not only the traditional Midas Touch problem but also enhance the overall user experience in virtual reality (VR) interactions.

Considering that visual depth is a new input dimension for VR users, we place emphasis on the learnability of our design. To evaluate this aspect, we conducted both a user study and a case study to observe how novice users adapt to this new interaction method. The results demonstrate that gaze-depth interaction can be learned smoothly with adaptive visual cues that facilitate users' perception of depth and develop muscle memory.

We hope this study will inspire more researchers to refine theoretical frameworks related to gaze-depth interaction. For instance, future work could focus on designing more effective learning processes, assisting users' depth perception during transition periods, and integrating inputs from other modalities. These efforts would provide valuable theoretical support for promoting widespread adoption of this intuitive interaction method.

%% file: tex/10-appendix.tex
\begin{figure*}[htb]
\centering
\includegraphics[width=0.8\linewidth]{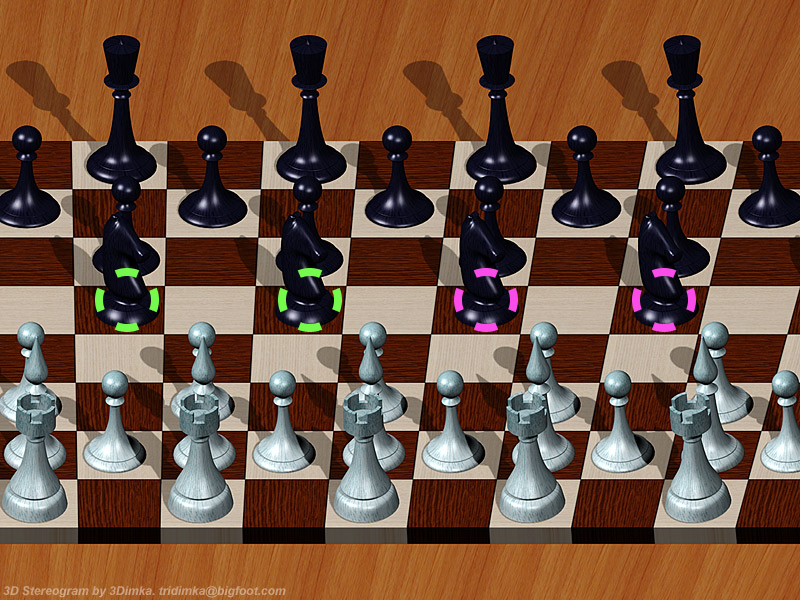}
\caption{Autostereogram Example. When you put your gaze point sightly further and make the green circle and pink circle overlap, you can see a 3D image. Making the image as large as possible and not too close to the image will help you see the 3D image successfully.}
\label{fig:Autostereogram}
\end{figure*}

\section{Autostereogram}
\label{Autostereogram}

Figure \ref{fig:Autostereogram} shows an \textit{autostereogram}, a picture can show 3D patterns at a certain visual depth.

\section{User Experience Questionnaire}
\label{Questionnaire}

\subsection*{1. Usability and Learnability}
\begin{itemize}
    \item The gaze-based interaction method was easy to understand.
    \item I quickly became comfortable using the gaze-based interaction method in VR.
    \item The gaze-based interaction method was intuitive, requiring little or no explanation.
\end{itemize}

\subsection*{2. Responsiveness and Efficiency}
\begin{itemize}
    \item The gaze-based interaction method was responsive to my eye movements.
    \item The gaze-based interaction method allowed me to perform "selection" operation efficiently.
    \item I felt in control while using the gaze-based interaction method to perform "selection" operation.
\end{itemize}

\subsection*{3. Comfort and Naturalness}
\begin{itemize}
    \item The gaze-based interaction method felt natural and immersive.
    \item I experienced minimal eye fatigue while using the gaze-based interaction method to perform "selection" operation.
\end{itemize}

\subsection*{4. Preference and Overall Experience}
\begin{itemize}
    \item I would prefer using the gaze-based interaction method over other VR interaction methods such as controllers.
    \item The gaze-based interaction method enhanced my overall VR experience.
\end{itemize}

\subsection*{5. Applicability in Various Contexts}
\begin{itemize}
    \item I believe the gaze-based interaction method could be useful in a variety of VR applications.
\end{itemize}